# Revisiting the Becker-Kistiakowsky-Wilson equation of state

Loann NERON [1,2,*] and Richard SAUREL [1,2]

[1] RS2N SAS, 371 chemin de Gaumin, 83640 Saint-Zacharie, France
[2] Aix Marseille Univ, CNRS, Centrale Marseille, LMA UMR 7031, Marseille, France
[*] Corresponding author: loann.neron@etu.univ-amu.fr, loann.neron@rs2n.eu

## Abstract

The Jones-Wilkins-Lee (JWL) equation of state (EOS) is the most popular equation of state used in hydrocodes to model detonation products thermodynamics resulting from high explosives. However, the JWL EOS presents several major difficulties. Namely, its range of validity and convexity is limited as its parameters are adjusted on a given reference curve. When used in conditions where the thermodynamic state varies significantly, computational failure may happen. It occurs frequently in multiphase flow computations when heat and mass transfers with other phases are present, as the resulting thermodynamic state is far from the reference curve used for adjustment. Moreover, the detonation products composition is absent in the JWL formulation. This is another important limitation when additional physics are involved, such as post-combustion, phase transition, or when dealing with non-ideal explosives. In this work, the Becker-Kistiakowsky-Wilson (BKW) EOS is considered instead. The BKW EOS is widely used in thermochemical codes, as parent EOS to provide the reference data used to fit JWL EOS parameters. The BKW EOS present several advantages. Its formulation considers variable gas species in the detonation products. No specific adjustment is needed for a given explosive, and only the detonation products composition requires computational efforts. Its range of validity is large, as the various parameters have been adjusted for each gas species separately. However, the BKW EOS also present difficulties which limit its application in hydrocodes. These difficulties are the aim of the present work. The BKW EOS is currently restricted to thermochemical codes as its mathematical formulation is computationally expensive. It also requires the gas chemical composition to be computed via an appropriate chemical equilibrium solver. In the present work, the computational cost of the BKW EOS is reduced thanks to the thermodynamic relaxation method of Neron et al. (2023). The variable composition of the detonation products is accounted for through specific relaxation terms computed once, with the help of conventional thermochemical codes. The condensed species, such as solid carbon, are modelled with the Cochran-Chan (CC) EOS, having a simpler thermodynamic formulation and being thermodynamically consistent compared to usual EOSs used for condensed species. This multiphase formulation made of BKW gas mixture with variable composition and condensed phases is considered in temperature and pressure equilibrium. It is embedded in a multiphase flow formulation, with the help of the thermodynamic relaxation method mentioned above. It shows comparable computational cost as the JWL formulation, with significantly enhanced physics capabilities and improved robustness.

# 1. Introduction

The Jones-Wilkins-Lee (JWL) [1] equation of state (EOS) is the most popular formulation used in hydrocodes dealing with the detonation of high explosives. The JWL parameters are adjusted along the isentrope emerging from the Chapman-Jouguet (CJ) point computed with a thermochemical code. Most thermochemical codes use the Becker-Kistiakowsky-Wilson (BKW) EOS as parent EOS for the CJ state computation as well as related isentrope, such as for example Cheetah [2], EXPLO5 [3], and DRAGON [4]. The JWL is consequently a reduced EOS compared to BKW. It is reduced in the sense that the gas composition is absent in the JWL formulation. Its parameters are only adjusted on the isentrope connecting the CJ point to lower pressure.

The JWL EOS is usually written under the Mie-Grüneisen (MG) form and seems to be the most used formulation in multidimensional hydrocodes. See for example [5,6,7] and many others. It is one of the most popular formulations as it offers a good compromise between computational cost and accuracy. However, the JWL EOS presents fundamental difficulties.

First, the JWL EOS range of validity is limited. JWL parameters are fitted along the isentrope emerging from the CJ point. However, MG-type EOS only tolerates small variations around their reference curve. This restricted domain of validity becomes problematic for multiphase computations where heat and mass exchanges between phases are present, resulting in significant thermodynamic state deviations. These restrictions lead to convexity issues, which appear as negative square sound speed or negative temperature, resulting in computational failure.

Moreover, the JWL EOS model the mixture of detonation products as a whole, i.e., all gas and condensed species are gathered under the same EOS and considered in thermodynamic equilibrium. This simplification is one of the main reasons for the JWL EOS popularity, since the complexity of the detonation products composition is implicitly considered during JWL parameters adjustment and hidden afterward. However, it also means that the composition, or related parameters, are only adjusted along the reference CJ isentrope. The detonation products composition cannot be evolved with the JWL EOS. This is a limitation when delayed effects, such as post-combustion or phase change, are present. And for non-ideal explosives where a significant part of the reaction occurs after the CJ point.

The present contribution proposes to directly use the parent BKW EOS to model detonation products instead of the JWL EOS. The BKW EOS present several key features which are improvements compared to the JWL EOS.

First, as most thermochemical codes use the BKW EOS to model detonation products, related thermodynamic and physical data have been carefully adjusted (see [8,9]), rendering the formulation independent to the explosive, contrarily to JWL.

Second, the BKW EOS considers gas species present in the detonation products. In the present work, each gas species considered with the BKW EOS has a mass fraction transport equation in the flow model. Source terms are added to vary the gas mixture composition according to local flow conditions. These sources terms are determined by following the same basic idea as for JWL. Here, the source terms are adjusted to agree with the flow composition variation along the isentrope emerging from the CJ point. Extra contributions may be added for post combustion effects or phase change. This modelling increases the range of validity compared to JWL. Moreover, since thermodynamic parameters for each species of BKW are calibrated over a wide range of experimental data, its validity and convexity range is also quite large.

However, the BKW EOS also has some limitations which usually prevent its use in hydrocodes.

The first one is its mathematical form, needing Newton loops for inversion to determine all thermodynamic variables. This complexity rises as the BKW EOS is only valid for gas mixtures, while detonation products contain condensed phases. Condensed species (solid carbon, liquid water, metals, unreacted explosive, ...) need their own equation of state. Thus, a mixture model is necessary for the BKW EOS and the condensed phase EOSs. Such mixture models are challenging to build and computationally expensive.

Additionally, in thermochemical codes, local thermodynamic equilibrium is computed at the CJ point and each point of the isentrope. In the frame of sophisticated EOS such as the BKW one, insertion of a thermochemical equilibrium solver in a hydrocode is possible but computationally very expensive.

To alleviate those difficulties, several key features are given in the present contribution.

- The thermodynamic relaxation method of Neron et al. [10] is adapted to the BKW EOS and used to speed up computations, specifically pressure relaxation solvers, mandatory for multiphase flows computations.
- Solid carbon and other possible condensed species are modelled with the Cochran and Chan [11] (CC) EOS under Mie-Gruneisen (MG) form. This EOS is simpler compared to the ones used in thermochemical codes such as Cheetah and appears nonetheless accurate enough. A similar approach was done by Suceska et al. [12] in the EXPLO5 code, where the condensed phases are considered through the Murnaghan EOS, a simplified MG formulation.
- The multiphase model of Saurel et al. [13] in pressure and temperature disequilibrium is used to compute two-phase mixtures, with gas detonation products governed by the BKW EOS and condensed phases governed by the CC EOS. This non-equilibrium approach appeared efficient to enforce pressure and temperature equilibrium among the phases. Mechanical and thermal relaxation are split in two steps, each one being simple and fast. It seemed more efficient than considering a single step with two constraints of pressure and temperature equilibrium, leading to a matrix Newton loop.
- No chemical equilibrium solver is used in the computations, as it would be too expensive. The composition variation is handled through source terms introduced into the mass and volume fraction equations of the flow model. These source terms are computed once, with the help of a thermochemical code, along the CJ isentrope.

In this work, only CHNO-type explosives are considered to simplify validations and highlight efficiency of the present modelling. Indeed, most explosives are of CHNO-type and many experimental data are available. Extension to other explosives such as aluminized ones should follow the same methodology.

The paper is divided into two main parts.

In the first part, only positive oxygen balance explosives are considered. These explosives do not produce condensed species such as solid carbon. This allows investigation of only the BKW EOS without additional ingredients necessary for condensed phases. The BKW EOS formulations and associated thermodynamic parameters are presented in Section 2. Then, the BKW EOS is implemented in the multi-constituents Euler equations in Section 3.1. Shock-tube tests for various explosives are done with constant composition and results are compared with predictions obtained with the JWL EOS in Section 3.2, showing good agreement and validating BKW as an alternative to JWL. A simple and novel method to account for changes in detonation product composition is developed in Section 3.3. Next, the capability of the BKW EOS to compute interface between gases is illustrated in Section 3.4. It is shown that the BKW EOS can capture interfaces without specific treatment, at least in explosion conditions. Lastly, the thermodynamic relaxation method of [10] is adapted to the BKW EOS and validated in the context of the Euler equations. This step is necessary to address multiphase extension, regarding computational saving.

In the second part, negative oxygen balance explosives are studied. Since only CHNO-type explosives are examined, solid carbon is the only condensed phase present. The CC EOS formulations and associated thermodynamic parameters for solid carbon are given in Section 4. Then, CJ conditions are computed for various explosives and initial densities in Section 5.1, using the BKW EOS to model gas species and the CC EOS for solid carbon. Predictions are compared to experimental data and computed data from the thermochemical code Cheetah. Shock-tube tests are also considered, and results are compared to JWL predictions in Section 5.2 showing again good agreement, validating the method to model condensed phases. In Section 5.3, mass transfer between the solid phase and the gas phase is developed to account for solid carbon variations. In Section 5.4, 1D spherical explosions of TNT are computed and results are compared to experimental pressure records. BKW-CC EOSs slightly improves predictions compared to JWL. Both thermodynamic formulations are also compared in terms of computational efficiency. Lastly, a simple post-combustion model is developed in Section 5.5 to show the BKW capabilities. Conclusions are given in Section 6.

## 2. Becker-Kistiakowsky-Wilson equation of state

### *2.1. Equation of state formulation*

The Becker-Kistiakowsky-Wilson (BKW) equation of state (EOS) is intensively used to calculate detonation properties in thermochemical codes (0D). Mader [14] and Suceska et al. [15] gives the historical background for the BKW EOS. The most known form nowadays was developed by Becker [16], Kistiakowsky and Wilson [17], Cowan and Fickett [18]. The BKW is usually expressed in molar quantities. However, mass quantities are more convenient for flow computations. Conversion of the molar expression in mass units is conventional and detailed for example in [19]. Thus, the BKW EOS is considered in this work under the form,

$$P(v,T,y_i) = \frac{RT}{v}(1 + Xe^{\beta X}), \qquad (1)$$

where $P$, $T$ and $v$ are respectively the pressure, the temperature and the specific volume of the gas mixture in the post-detonation state. The specific volume corresponds to the inverse of the density $\rho$. $R$ is the specific gas constant of the mixture,

$$R = \hat{R} \sum \frac{y_i}{W_i}. \qquad (2)$$

$\hat{R} = 8.314$ J/mol/K is the universal gas constant. $y_i$ and $W_i$ corresponds respectively to the mass fraction and molar mass of species $i$. The saturation constraint implies $\sum y_i = 1$. The term $X$ is a function depending on temperature, specific volume and mass fractions. It reads,

$$X = \frac{\kappa \sum \frac{y_i}{W_i} k_i}{v(T+\theta)^\alpha}. \qquad (3)$$

Parameters $\alpha$, $\beta$, $\theta$ and $\kappa$ are constant coefficients. $k_i$ is the covolume of species $i$.

Equation (1) corresponds to the thermal BKW EOS. For application in hydrodynamic codes, knowledge of the associated caloric EOS for the internal energy $e = e(v,T,y_i)$ is needed. Thermal and caloric EOS must fulfil a compatibility condition derived from Maxwell rules (see for example [20]). The caloric EOS compatible with (1) is derived from the compatibility condition,

$$\left.\frac{\partial e}{\partial v}\right|_{T,y_i} = T\left.\frac{\partial P}{\partial T}\right|_{v,y_i} - P. \tag{4}$$

The partial derivative of the pressure according to the temperature is,

$$\left.\frac{\partial P}{\partial T}\right|_{v,y_i} = \frac{R}{v}\left(1 + \left(1 - \frac{\alpha T}{T+\theta}(1+\beta X)\right)Xe^{\beta X}\right). \tag{5}$$

The pressure (1) and derivative (5) are introduced in the compatibility condition (4),

$$\left.\frac{\partial e}{\partial v}\right|_{T,y_i} = -\frac{\alpha R T^2 X (1+\beta X) e^{\beta X}}{v(T+\theta)}. \tag{6}$$

Relation (6) is integrated with respect to the specific volume, yielding the caloric BKW EOS,

$$e(v,T,y_i) = \frac{\alpha R T^2 X e^{\beta X}}{T+\theta} + e_{IG}(T,y_i). \tag{7}$$

With $e_{IG}(T,y_i)$ a function of the temperature and mass fractions. This function represents the ideal gas (IG) contribution to the internal energy and is expressed as,

$$e_{IG}(T,y_i) = \sum \frac{y_i}{W_i}\left(a_i T^2 + b_i T + c_i\right). \tag{8}$$

Parameters $a_i$, $b_i$ and $c_i$ are constant coefficients specific to each gas species $i$. These coefficients are adjusted according to the enthalpy of formation of the considered species. More details are given in Section 2.2.

In compressible fluids, knowledge of the speed of sound is fundamental. The square sound speed is defined as,

$$c^2(v,T,y_i) = \left.\frac{\partial P}{\partial \rho}\right|_{s,y_i}, \tag{9}$$

where $s$ is the entropy. However, determination of the entropy in terms of pressure and density for the BKW EOS is quite challenging. After some manipulation, a more practical definition of the square sound speed is obtained,

$$c^2(v,T,y_i) = -v^2\left(\left.\frac{\partial P}{\partial v}\right|_{T,y_i} - \frac{T}{\left.\frac{\partial e}{\partial T}\right|_{v,y_i}}\left(\left.\frac{\partial P}{\partial T}\right|_{v,y_i}\right)^2\right). \tag{10}$$

The partial derivative of the pressure according to the specific volume is,

$$\left.\frac{\partial P}{\partial v}\right|_{T,y_i} = -\frac{RT}{v^2}\left((2+\beta X)Xe^{\beta X} + 1\right). \tag{11}$$

The partial derivative of the energy according to the temperature, which is by definition the specific heat at constant volume $C_v$ of the BKW EOS, reads,

$$\left.\frac{\partial e}{\partial T}\right|_{v,y_i} = C_v = \frac{\alpha RT X e^{\beta X}}{(T+\theta)^2}\left(2\theta + T - \alpha T(1+\beta X)\right) + \sum \frac{y_i}{W_i}(2a_i T + b_i). \tag{12}$$

Having in hands (5), (11) and (12), the square sound speed (10) is available.

Convexity of the BKW EOS is addressed in Appendix A.

Now that all needed thermodynamic functions for the BKW EOS are available, determination of the various thermodynamic parameters is addressed.

## *2.2. Thermodynamic parameters*

In this section, parameters of the BKW EOS used throughout this work are detailed. These parameters are divided into two categories. The main parameters are $\alpha$, $\beta$, $\theta$, $\kappa$ and $k_i$, for which many sets of data are available in the literature. The other parameters are related to the ideal gas part of the internal energy. Parameters $a_i$, $b_i$ and $c_i$ are adjusted for each species considered.

Many sets of the parameters $\alpha$, $\beta$, $\theta$, $\kappa$ and $k_i$ are available in the literature. For example, [14] proposed two sets of parameters. BKW-TNT adjusted with detonation velocities data from TNT, for high density explosives containing large amount of solid carbon in the detonation products. And BKW-RDX adjusted for explosives with low amount of solid carbon in the same way. To improve detonation velocity predictions, [21] proposed the BKWR set of parameters, [8] the BKWS set, and [22] the BKWC set. More details are available in [23] and [9] which also proposes other sets of parameters. Each new set of parameters aim to better predict experimental detonation data. However, it proves quite challenging to find a universal set of parameters. In the present work, the BKWC set of parameters is arbitrarily chosen as it is the most recent one in the Cheetah 2.0 code [2] used by the authors as thermochemical solver. The BKWC EOS thermodynamic parameters are summarized in Table 1.

| Parameter | | Value |
|---|---|---|
| $\alpha$ | | 0.5 |
| $\beta$ | | 0.403 |
| $\theta$ (K) | | 5441 |
| $\kappa$ (m$^3$.K$^\alpha$/mol) | | $10.86 \times 10^{-6}$ |
| $k_i$ | $CO_2$ | 511 |
| | $H_2O$ | 188 |
| | $N_2$ | 374 |
| | $O_2$ | 306 |
| | CO | 372 |
| | NO | 394 |

Table 1. BKW EOS thermodynamic parameters and covolumes corresponding to the BKWC set of parameters given by [22].

Only the six gas species present in Table 1 are considered in this work since they are the main species present in detonation products for CHNO explosives. Indeed, the goal is to have an accurate but fast thermodynamic formulation, so only the main species are retained, since the other species considered in BKWC computations are always in minor concentration for the explosives studied.

The ideal gas energy contribution $e_{IG}$ parameters $a_i$, $b_i$ and $c_i$ need to be determined for each species. These coefficients are adjusted using the energy of formation for the considered species. A root-mean-square fitting method is used with an added condition that the function is equal to the energy of formation in the standard state $P_0 = 1$ bar and $T_0 = 298.15$ K. The data from the JANAF tables [24] are used to fit the parameters. For each species, JANAF thermochemical tables gives $\Delta \hat{H}_f^0$, the difference between the enthalpy of formation at a given temperature and at the standard temperature. The ideal gas contribution used to adjust the parameters is obtained from the tables as,

$$e_{IG}(T) = \frac{\Delta \hat{H}_f^0 + \hat{H}_f^0(T_0) - \hat{R}T}{W}. \tag{13}$$

The coefficients obtained for each species are summarized in Table 2, and corresponding fits are shown in Figure 1.

| Species | $W_i$ (g/mol) | $a_i$ (J/mol/K$^2$) | $b_i$ (J/mol/K) | $c_i$ (J/mol) |
|---|---|---|---|---|
| CO$_2$ | 44 | 0.00155347 | 42.7380 | -408881 |
| H$_2$O | 18 | 0.00238886 | 30.3603 | -253569 |
| N$_2$ | 28 | 0.00072051 | 23.4866 | -9545 |
| O$_2$ | 32 | 0.00108440 | 24.6865 | 80397 |
| CO | 28 | 0.00066576 | 24.0172 | -120229 |
| NO | 30 | 0.00061462 | 24.3285 | -9829 |

Table 2. Ideal gas contribution $e_{IG}$ coefficients and molar mass $W$ for the various species considered in the detonation products with BKW EOS.

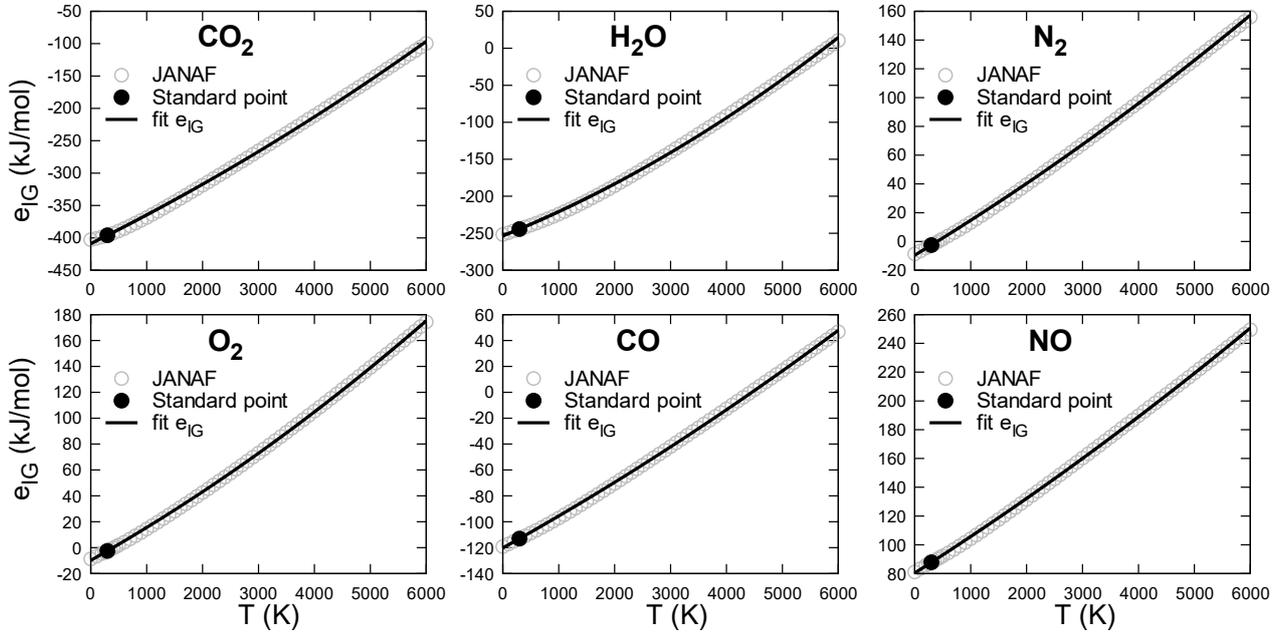

Figure 1. Ideal gas contribution depending on the temperature for the various gas species considered in BKW EOS. Grey circles represent experimental data from JANAF thermochemical tables [24] with the standard point at $T_0 = 298.15$ K filled with black. The fitted ideal gas contribution function $e_{IG}$ is represented with the black line and shows good agreement with experimental data.

In the literature, the caloric BKW EOS is scarcely detailed or absent. There seems to be no consensus on how to handle the ideal gas contribution to the energy. [25] models this contribution as the energy obtained from a fit of the entropy with a fourth order polynomial function. In the thermochemical code CHEMKIN [19,26], the energy is computed from the specific heat at constant pressure $C_p$ fitted to experimental data. In the thermochemical code Cheetah [27], a similar method is used.

In the stated examples of the literature and for the method used in this work, the idea is essentially the same. A temperature dependent function is fitted to experimental data, i.e., the JANAF thermochemical tables, and the ideal gas contribution to the energy is computed from the fitted function. The function (8) used herein is rather simple with three coefficients only. Nonetheless, the agreement with experimental data appears satisfying (Figure 1), at least for the considered species.

The BKW EOS formulations being now available, the EOS is tested in the context of Euler equations in the next sections.

## 3. BKW without condensed carbon

The BKW EOS addresses only gas species present in the detonation products. For CHNO explosives with negative oxygen balance, which corresponds to most explosives, condensed carbon is also present in the detonation products. Condensed carbon consideration adds an extra level of complexity and will be investigated further in Sections 4 and 5. This section focusses solely on the BKW EOS, embedded in the Euler equations with mass fraction equations. Only positive oxygen balance explosives (Ammonium Nitrate (AN), Nitroglycerin (NG) and Tetranitromethane (TNM)) are considered since their decomposition does not produce solid carbon or other condensed species. Predictions obtained with the BKW EOS are compared to results with the JWL EOS for 1D shock-tube tests in Section 3.2. The first tests are achieved with constant composition. The impact of the detonation products composition variation is investigated in Section 3.3 with the help of an instantaneous relaxation method. Then, the capability of the BKW to treat interface with air is illustrated in Section 3.4. Last, the thermodynamic relaxation method of [10] is adapted to the BKW EOS in Section 3.5, to speed up computations in the direction of multiphase flow models, needed for condensed phases consideration.

### 3.1. Euler equations with mass fractions

The 1D Euler equations of gas dynamics are recalled hereafter. Transport equations for mass fractions $y_i$ of the various gas species considered with the BKW EOS are added.

$$\begin{cases} \dfrac{\partial \rho}{\partial t} + \dfrac{\partial \rho u}{\partial x} = 0, \\ \dfrac{\partial \rho u}{\partial t} + \dfrac{\partial \left( \rho u^2 + P \right)}{\partial x} = 0, \\ \dfrac{\partial \rho E}{\partial t} + \dfrac{\partial \left( \rho E + P \right) u}{\partial x} = 0, \\ \dfrac{\partial \rho y_i}{\partial t} + \dfrac{\partial \rho y_i u}{\partial x} = 0, \quad \forall i. \end{cases} \quad (14)$$

Most notations have been defined previously. The fluid velocity is denoted by $u$ and the total energy by $E = e + u^2/2$. Six mass fractions equations, corresponding to the six chemical species of Table 2, are considered.

These equations with the BKW EOS allow non-conventional capabilities. For example, a given zone of the domain can be initialized with detonation products composition, while the rest of the domain can be initialized with air, creating an interface between air and detonation products. This capability may help for explosions in air. Usually, specific methods (diffuse interfaces, VOF, Front Tracking) may be used to deal with mixture cells. Obviously, as the present flow model involves a single temperature, perfect matching of fluids velocities and pressure among the fluids may be lost. This will be examined in Section 3.4.

Another interesting example with System (14) and BKW EOS is the modification of the detonation products composition caused by secondary reactions such as post-combustion. In this aim, source terms are needed in the mass species equations. This flexibility is unavailable with the JWL EOS since the composition is hidden in the EOS parameters and cannot be modified afterward.

However, as stated in Introduction, even if the BKW EOS is widely used in thermochemical codes, it is hardly used in flow codes due to its complicated formulation. One difficulty is that the key thermodynamic function linking the pressure and the internal energy $P = P(v,e)$ is not explicitly available for the BKW EOS. An iterative method is needed to solve this function which can be needed several times per time step depending on the flow model complexity. This adds computational cost when BKW is used. The additional cost remains acceptable in the present context of the Euler equations. However, for negative oxygen balance explosive, multiphase flow models are needed to account for condensed phases. The computational cost dramatically increases with BKW, mainly because of pressure relaxation solvers. The added computational cost will be reduced thanks to the thermodynamic relaxation method of [10].

In the following test cases, the flow model (14) is solved numerically with a Godunov-type method using the HLLC approximate Riemann solver to compute the intercell flux (see [28,29]). All computations are first-order accurate in both space and time, to avoid any additional numerical artifact. The time step is computed with CFL (Courant-Friedrichs-Lewy) number of 0.8. The computational domain is 1-m long and is spatially resolved with 1000 cells. For each test case, the results obtained with the BKW EOS are compared with the predictions obtained with the JWL EOS. Formulation of the JWL EOS and associated thermodynamic parameters are given in Appendix B. The main objective is to ensure that computations with the BKW EOS run correctly and predicts results similar to the JWL EOS. Indeed, JWL being the reduced form of BKW used in most hydrocodes. Shock-tube tests for each explosive having positive oxygen balance are considered.

### 3.2. Shock-tube tests

1D shock-tube tests are computed for each positive oxygen balance explosive (AN, NG and TNM) considered. For all explosives, the high-pressure (HP) left chamber is initially set in the corresponding CJ conditions $(P_{CJ}, \rho_{CJ})$. The low-pressure (LP) right chamber is initially set to the density $\rho_0/10$ and the corresponding pressure along the isentrope from the CJ point, with $\rho_0$ the initial density of the explosive material. Both points are considered for fitting JWL parameters, so it ensures that the JWL EOS is not used outside its range of validity during the computations. Data for both the CJ point and low-pressure point along the isentrope are obtained with Cheetah and are summarized in Table 3. Initial compositions for the BKW EOS mass fractions $y_i$ in each zone are also computed with Cheetah using the BKWC library.

|  | | AN | | NG | | TNM | |
|---|---|---|---|---|---|---|---|
| Variable | | HP | LP | HP | LP | HP | LP |
| $P$ (bar) | | 46600 | 178 | 223800 | 924 | 137500 | 260 |
| $\rho$ (kg/m$^3$) | | 1375 | 105 | 2094 | 160 | 2085 | 164 |
| $y_i$ | $CO_2$ | 0 | 0 | 0.566 | 0.581 | 0.222 | 0.225 |
| | $H_2O$ | 0.450 | 0.450 | 0.198 | 0.198 | 0 | 0 |
| | $N_2$ | 0.350 | 0.350 | 0.180 | 0.185 | 0.289 | 0.285 |
| | $O_2$ | 0.200 | 0.200 | 0.036 | 0.035 | 0.481 | 0.490 |
| | CO | 0 | 0 | 0.010 | 0 | 0 | 0 |
| | NO | 0 | 0 | 0.010 | 0.001 | 0.008 | 0 |

Table 3. Initial data of the 1D shock-tube tests for the various positive oxygen balance explosives considered. All values are obtained with the thermochemical code Cheetah using the BKWC library.

The initial discontinuity is located at 0.6 m. The computed results for AN, NG and TNM are shown in Figure 2 with each column corresponding to each explosive respectively. Each row corresponds to computed flow results: pressure, velocity, density, and temperature.

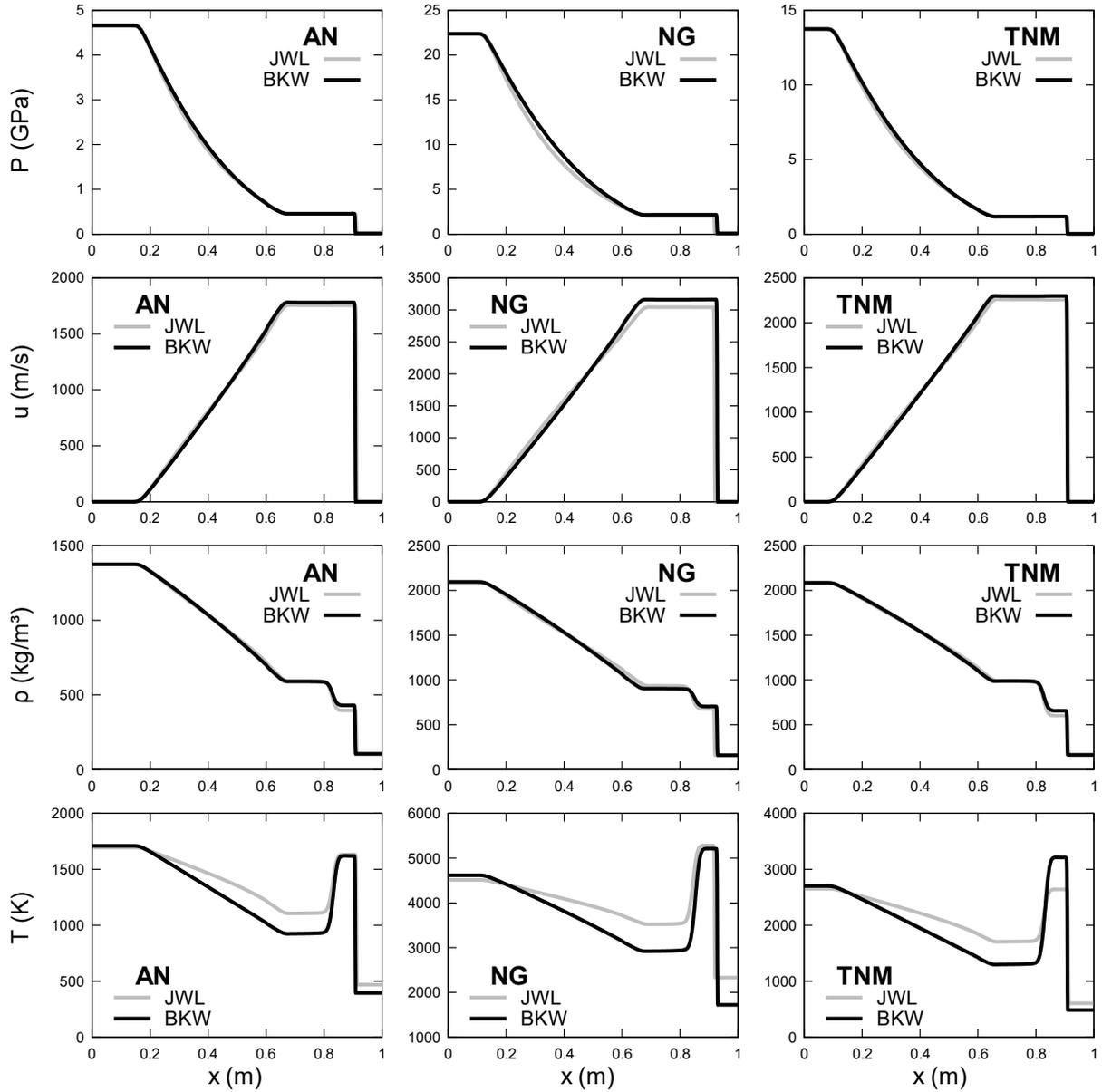

Figure 2. Shock-tube test for the AN, NG and TNM explosives. Initial states data are given in Table 3. The initial discontinuity is located at 0.6 m. Results are shown at time 130 µs, 80 µs and 100 µs respectively. Grey lines represent predictions with the JWL EOS, and black lines with the BKW EOS.

For all considered explosives, predictions with the JWL EOS and BKW EOS are quite close for the pressure, density and velocity. Density and velocity plateau values are a bit higher with the BKW EOS. These discrepancies remain minor and are expected, since there is no reason for both EOS to predict merged results. At this point, the BKW EOS seems capable of predictions similar to the JWL EOS without any particular issue.

The main difference between both EOS is related to temperatures. Indeed, for all explosives, except initially in the CJ state, predicted temperature with BKW and JWL are different, especially in expansion waves behind the shock with up to 500 K differences. This deviation can be explained by two reasons. First, during the JWL EOS parameters adjustment, no effort is done to satisfy the temperature along the isentrope except at the CJ point. Second, the detonation products composition with BKW EOS remains constant. Indeed, with the flow model (14), mass fractions are only transported in the flow. However, depending on the flow conditions, the composition may vary. For example, before and after the left-propagating expansion wave, the density is divided by two. In the present tests, the composition used in the BKW EOS in this zone remains the same, namely the composition at CJ state. Since the JWL EOS is fitted using several points along the isentrope from the CJ point, the parameters implicitly consider the variation of composition in the detonation products. The impact of the composition variation on BKW EOS predictions is investigated in the next Section 3.3. The validity of JWL temperature predictions is also investigated.

### *3.3. Variation of the chemical composition*

The detonation products composition varies depending on the thermodynamic conditions. Usually, the BKW EOS is used in thermochemical codes, with a chemical equilibrium solver to compute the detonation products composition. However, such equilibrium solvers are computationally too expensive to be used in flow solvers. In the present work, a simplified method is considered to account for changes in the composition. Source terms are added in the mass fractions equations of the flow model (14),

$$\frac{\partial \rho y_i}{\partial t} + \frac{\partial \rho y_i u}{\partial x} = \rho \dot{y}_i = \frac{\rho}{\tau}\left(y_i^*(\rho) - y_i\right), \quad \forall i, \text{ with } \tau \to 0^+. \tag{15}$$

The mass transfer is supposed instantaneous as $\tau \to 0^+$. It means that, after the hyperbolic step, instantaneous relaxation for each gas species is achieved. The updated mass fraction of species $i$ is given by the function $y_i^*(\rho)$. This density dependent function is a linear approximation of the variation of each species mass fraction between the CJ point and atmospheric conditions. Instead of computing the chemical equilibrium with a thermochemical code, the mass fractions are supposed to evolve linearly between two given points denoted "1" (CJ) and "2" (atmospheric). The function is just a linear interpolation,

$$y_i^*(\rho) = y_{i,1} + \left(y_{i,2} - y_{i,1}\right)\frac{\rho - \rho_1}{\rho_2 - \rho_1}. \tag{16}$$

An important remark regarding the approximation (16) is that, assuming $\sum y_{i,1} = 1$ and $\sum y_{i,2} = 1$, the new composition obtained after relaxation inherently satisfies the saturation constraint $\sum y_i^* = 1$. This is because the term $\frac{\rho - \rho_1}{\rho_2 - \rho_1}$ is constant for all species.

The linear approximation for mass fraction variation with density appears to be sufficient, at least for the positive oxygen balance explosives considered here. For AN, the composition is nearly invariant between the CJ state and atmospheric state. For NG and TNM, some species mass fractions remain constant, while others vary slightly and monotonously. Let's consider for example the NG explosive. The detonation products compositions are computed with Cheetah for several points along the isentropic expansion starting from the CJ point. The mass fractions for each species and densities in CJ conditions and atmospheric conditions used in the function (16) are summarized in Table 4. The mass fractions for the different gas species as a function of density are shown in Figure 3.

| Parameter | | CJ | Atmospheric |
|---|---|---|---|
| $\rho \, (\text{kg/m}^3)$ | | 2094 | 1.28 |
| $y_i$ | $CO_2$ | 0.565404 | 0.581477 |
| | $H_2O$ | 0.198155 | 0.198261 |
| | $N_2$ | 0.180377 | 0.185016 |
| | $O_2$ | 0.035740 | 0.035246 |
| | $CO$ | 0.010197 | 0 |
| | $NO$ | 0.010127 | 0 |

Table 4. Density and mass fraction of the gas species used in the BKW EOS for NG at the CJ point and atmospheric point. Data are computed with Cheetah using the BKWC library.

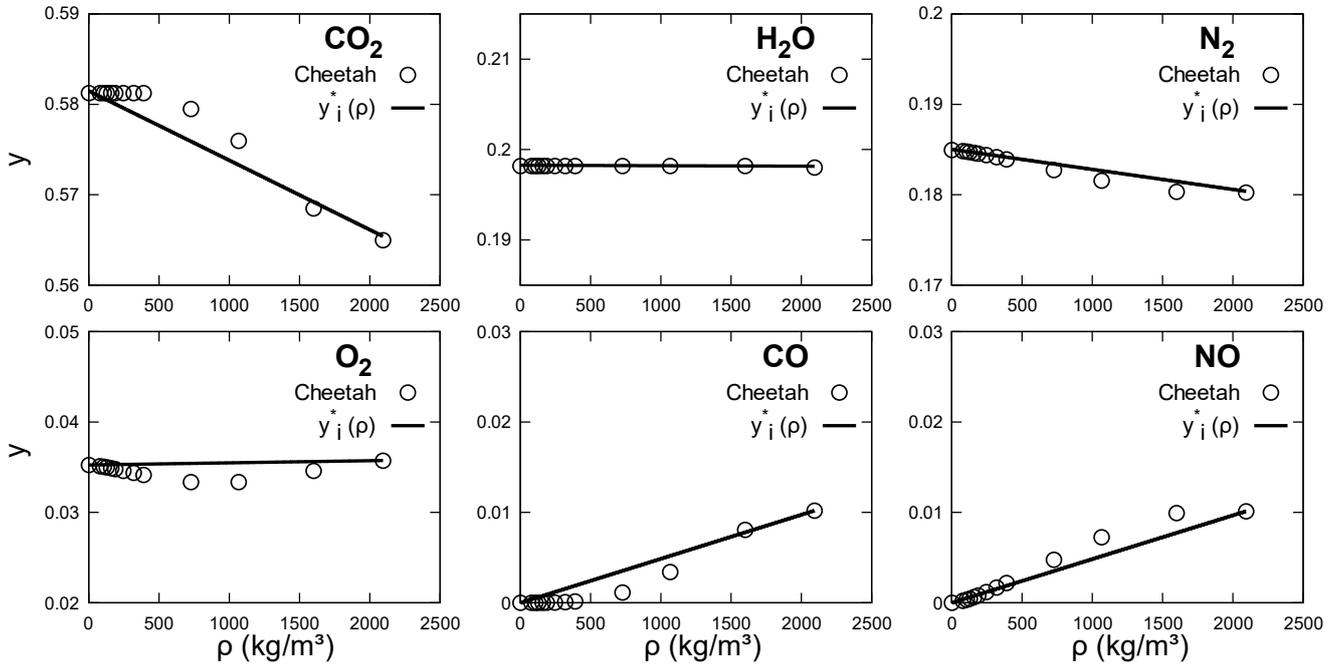

Figure 3. Mass fraction of gas species for NG detonation products along the isentropic expansion starting from the CJ point as a function of density. Reference points computed with Cheetah using the BKWC library are indicated with black circles. The linearly approximated mass fractions with relation (16) are represented in black lines. The linear approximation seems accurate enough.

From Figure 3, the linear approximation of the mass fraction appears accurate enough to represent the composition variations, at least along the isentropic expansion from the CJ point. A second-order approximation or higher could be used to improve this simplified method. However, as will be illustrated hereafter, the composition only slightly varies and the impact on BKW predictions is negligible. For negative oxygen balance, when solid carbon is present, this model is insufficient. Solid carbon will be governed by its own phase and a mass exchange model between phases will be developed. This point will be investigated in Section 5.3.

The impact of the detonation products composition variation is examined in the previous shock-tube tests presented in Figure 2. Only the NG explosive is shown since results and conclusions are similar for the other explosives. Mass fractions are initialized with relation (16). Results are shown in Figure 4 for the JWL EOS and the BKW EOS with and without variation of the composition. The corresponding mass fractions of each species are presented in Figure 5.

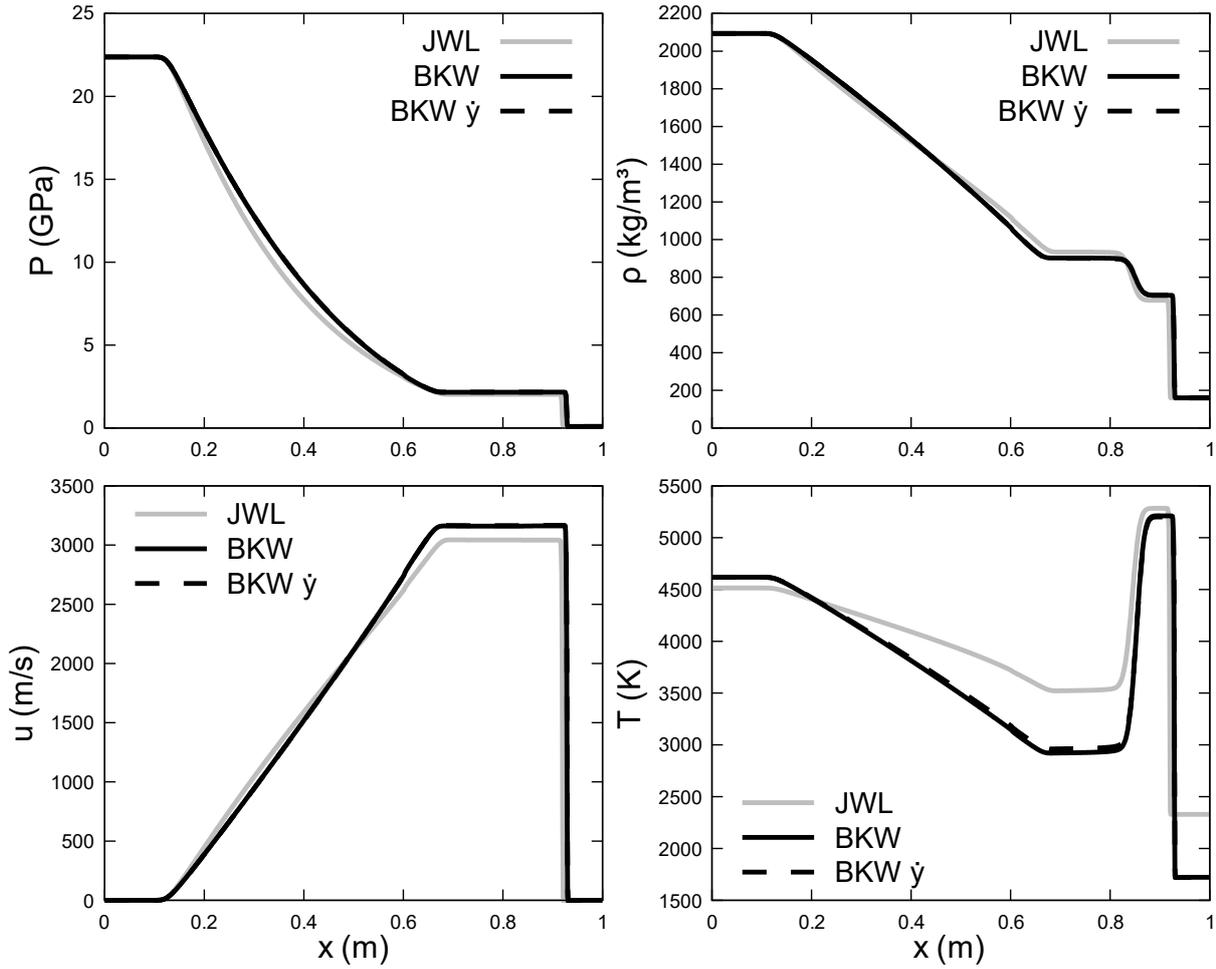

Figure 4. Shock-tube test for the explosive NG. The initial discontinuity is located at 0.6 m. Results are shown at time 80 µs. Grey lines represent results with the JWL EOS. Black lines represent predictions with the BKW EOS with invariant composition, and dashed lines with variable composition. Results of both BKW computations are merged except for the temperature where slight differences appear.

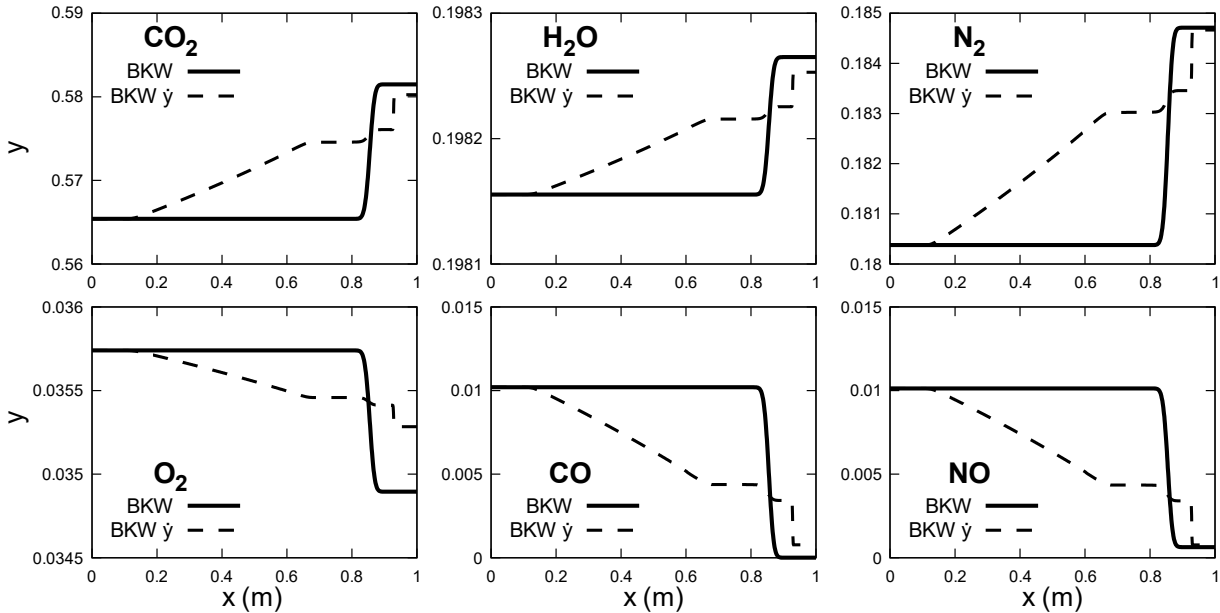

Figure 5. Mass fraction of each gas species of the shock-tube test for the explosive NG related to Figure 4. Black lines represent predictions with the BKW EOS with invariant composition, and dashed lines with variable composition.

The impact of the composition variation in this example is negligible on the thermodynamic variables shown in Figure 4. This is not surprising since between the CJ point and the atmospheric point, the NG detonation products composition varies only slightly according to Cheetah results, reported in Table 4. This is also the case for the TNM and AN explosives. The impact of the varying composition on the shock-tube tests studied previously in Section 3.2 is similarly negligible. Thus, the obtained results are not shown here.

The conclusions are the same as for previous shock-tube tests, the JWL and BKW EOS predictions are close. The difference in temperature between the BKW EOS and JWL EOS observed previously is still apparent. The variable composition increases slightly the predicted temperature with BKW but is not enough to explain this difference. The main explanation is that the JWL EOS is inaccurate for temperature calculations.

As stated earlier, during the JWL EOS parameters adjustment, no specific effort is done to match the temperature except at the CJ point. JWL thermodynamic parameters are adjusted to fit the isentropic expansion from the CJ point in the $(v, P)$ plane computed with Cheetah using the BKW EOS. The temperature along the isentrope is not considered beside the CJ point temperature. Contrarily, the BKW EOS is certainly more accurate for temperature computation, as the ideal gas energy contribution (8) has been fitted for each species over a large temperature range. To assess the temperature prediction accuracy for both EOS, temperatures along the isentropic expansion starting from the CJ point are computed with both thermodynamic formulations. Pressure and density for various points are computed with Cheetah and the temperature is calculated with each EOS using $T(v, P)$. For the BKW EOS, two situations are considered. In the first one, the CJ state composition is imposed constant everywhere. In the second situation, the composition varies according to relation (16). Results are compared to Cheetah predictions for all explosives in Figure 6.

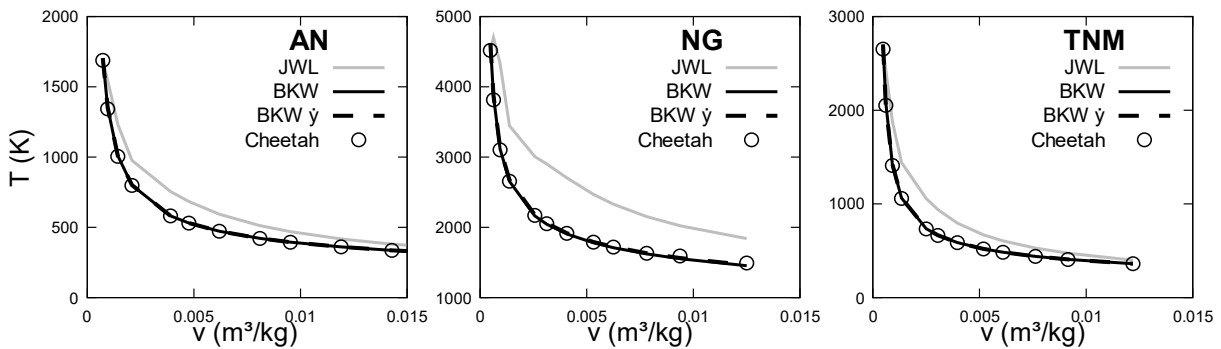

Figure 6. Temperatures along the adiabatic expansion starting from the CJ point in terms of specific volume. Reference points computed with Cheetah using the BKWC library are indicated with circles. Grey lines represent results with the JWL EOS. Black lines represent predictions with the BKW EOS with constant composition, and dashed lines with variable composition. The BKW EOS recovers Cheetah predictions, while the JWL EOS overestimate the temperature except at the CJ point (first point). The composition variation has no impact and both curves for BKW are merged.

The temperature predicted with the JWL EOS is overestimated compared to Cheetah computations, except at the CJ point which is specifically recovered during parameters adjustment. On the other hand, the BKW EOS results are in agreement with Cheetah computations. This is not surprising as this thermochemical code uses the same BKW EOS. Results are nonetheless slightly different since the ideal gas energy contribution formulation is different, as mentioned in Section 2.2. This temperature deviation explains the differences observed between both EOS in preceding shock-tube tests. Therefore, the JWL EOS seems inaccurate for temperature computation, while the BKW EOS seems more appropriate.

Moreover, the impact of the composition variation also appears negligible here. Consequently, for the rest of the numerical test, the composition of the detonation products will be considered constant (i.e., $\tau \to +\infty$ in (15)) unless stated otherwise.

The next Section 3.4 presents another interesting feature of the present formulation, namely modelling material interface with the same BKW EOS.

### *3.4. Mass fraction discontinuity: interface with air*

Another nice feature of System (14) with BKW EOS is the ability to model different fluids with only one BKW EOS by changing the mass fractions accordingly. A direct application is to model air, which is useful for example to study explosions in this medium. Representing air with the BKW EOS is easily done by initializing the mass fraction of the species $N_2$ and $O_2$ to match air composition: $y_{N_2} = 0.74$ and $y_{O_2} = 0.26$. Consequently, the BKW EOS allows the simulation of a multiphase situation only with the Euler equations and mass fraction transport equations of System (14). To model a similar situation with the JWL EOS, a more sophisticated multiphase flow model is needed. An option is to use the diffuse interface model of [13] as detailed in Appendix C. This option was used in this specific context in [30]. Apart from the obvious difficulty to implement such a multiphase model compared to the flow model (14), the computational cost is also increased, mainly because of pressure relaxation solvers.

As an illustration, shock-tube test cases similar to those of Section 3.2 are computed with air at atmospheric condition in the low-pressure chamber ($P = 1$ bar and $\rho = 1.17$ kg/m$^3$). Predictions with either the BKW or JWL EOS to model detonation products are compared. When the BKW EOS is used, the flow model (14) is solved. The air is governed by the same BKW EOS and the composition in the low-pressure chamber is initially set to $y_{N_2} = 0.74$ and $y_{O_2} = 0.26$. When the JWL EOS is used, the flow model presented in Appendix C is solved. The air is governed by the ideal gas EOS with $\gamma = 1.4$ and $C_v = 719$ J/kg/K. The two phases are in mechanical equilibrium (pressure and velocity). The initial discontinuity is located at 0.5 m. Results are shown in Figure 7 only for the NG explosive since results and conclusions are similar for the other explosives.

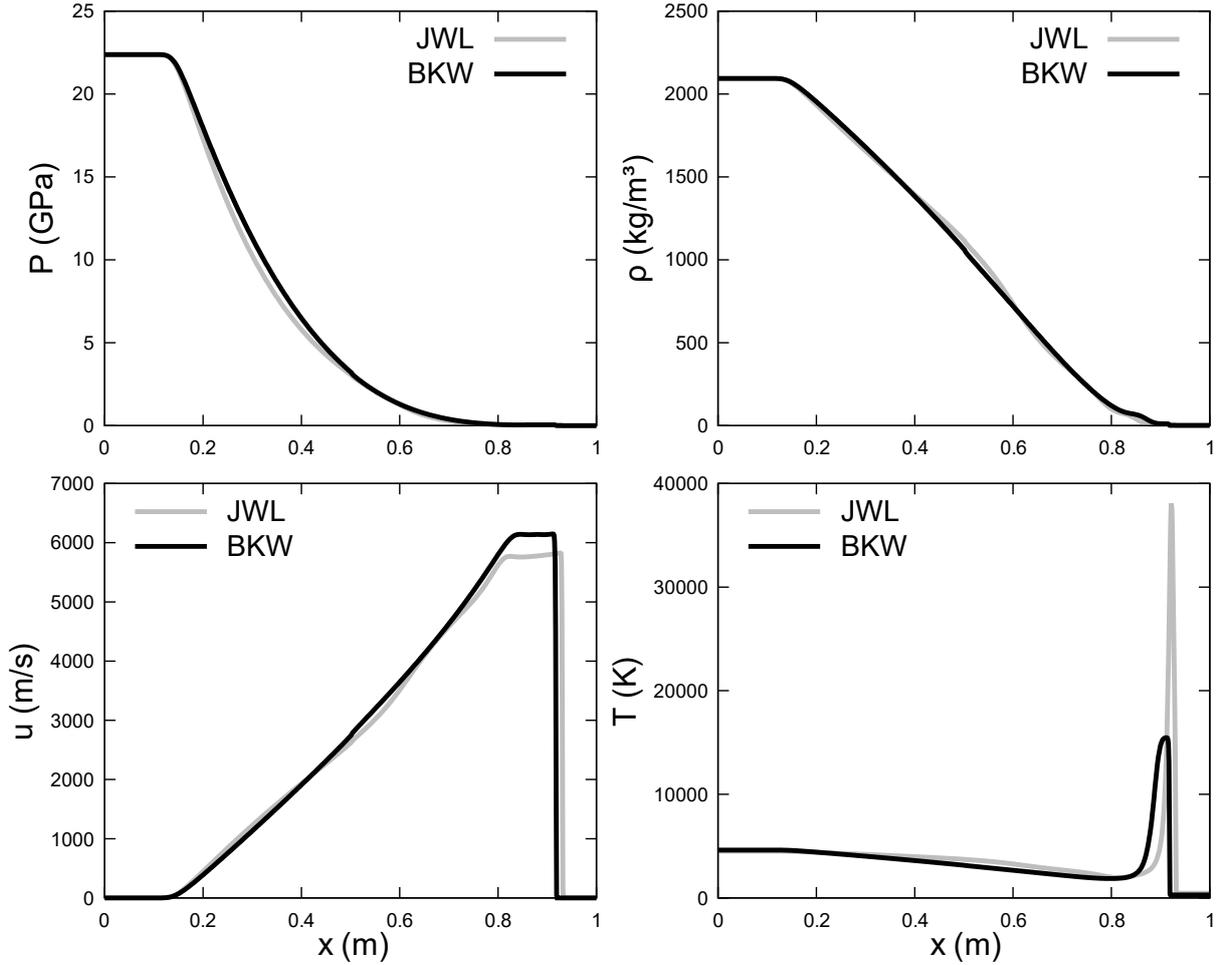

Figure 7. Shock-tube test for the NG explosive in air. The left state corresponds to the detonation products in CJ conditions and the right state corresponds to the air. The initial discontinuity is located at 0.5 m. Results are shown at time 60 µs. Grey lines represent predictions with the JWL EOS and the flow model of [13]. Black lines represent results with the BKW EOS and the flow model (14).

Results with the BKW method and the JWL method are again rather similar, with the same difference on the temperature as before. Discrepancies at the interface between detonation products and air are amplified. This was expected as the two models cannot be strictly compared. With the BKW EOS, the mixture is in temperature and pressure equilibrium. With the multiphase model, the detonation product and air are only in pressure equilibrium. The temperatures are different among the phases, which has some impact at the interface.

Regarding computational time, for all the previous tests done with the Euler equations, computations based on the JWL EOS are approximately two times faster than computations done with the BKW EOS. Contrarily, in the present example, the JWL computations are approximately 6 times slower compared to BKW. Indeed, the multiphase model with pressure relaxation needed to model air and detonation products is more expensive than the Euler equations.

In the present applications with the BKW EOS and the Euler equations (14), computational cost is reasonable. Approximately 1 second for the various tests is needed. However, with the consideration of solid carbon through an additional equation of state, the use of a multiphase flow model is necessary. Such flow models are more complex to solve (see Appendix C), and computational time increases dramatically when the BKW EOS is used. To improve the reliability of the BKW EOS in terms of computational performances, the thermodynamic relaxation method of [10] is adapted to the BKW formulation. The main ingredients of the method and its adaptation to the BKW EOS are first investigated in the context of the flow System (14) for the sake of simplicity. This is the aim of the following Section 3.5.

*3.5. Thermodynamic relaxation method for the BKW EOS*

Neron et al. [10] proposed a thermodynamic relaxation method to compute flows with sophisticated EOS such as Mie-Grüneisen EOS. The aim was to speed up and render robust computations. In this section, the same direction is followed, and the method is adapted to the BKW EOS. The method is presented within the framework of the flow model (14) to first validate it in a simple context. No computational gain with the thermodynamic relaxation method is expected for this flow model. But it is a necessary step before moving to multiphase flow computations with solid carbon.

The thermodynamic relaxation method consists in two steps. First, a predictor EOS (PEOS) under the form Noble-Abel-Stiffened-Gas (NASG) [31] is used during the hyperbolic step. The NASG EOS formulation reads,

$$\begin{cases} P(v,e) = \dfrac{R^*(e-q^*)}{C_v^*(v-b^*)} - \left(\dfrac{R^*}{C_v^*}+1\right)P_\infty^*, \\ P(v,T) = \dfrac{R^*T}{v-b^*} - P_\infty^*, \\ e(v,T) = C_v^*T + P_\infty^*(v-b^*) + q^*, \\ c^2(v,P) = \dfrac{v^2(R^*+C_v^*)(P+P_\infty^*)}{C_v^*(v-b^*)}. \end{cases} \quad (17)$$

The thermodynamic parameters $b^*$, $R^*$, $C_v^*$, $P_\infty^*$ and $q^*$ of the PEOS are transported in the flow with additional transport equations under conservative form,

$$\begin{cases} \dfrac{\partial \rho b^*}{\partial t} + \dfrac{\partial \rho b^* u}{\partial x} = \dfrac{\rho}{\tau}\left(b(v,e)-b^*\right), \\ \dfrac{\partial \rho P_\infty^*}{\partial t} + \dfrac{\partial \rho P_\infty^* u}{\partial x} = \dfrac{\rho}{\tau}\left(P_\infty(v,e)-P_\infty^*\right), \\ \dfrac{\partial \rho R^*}{\partial t} + \dfrac{\partial \rho R^* u}{\partial x} = \dfrac{\rho}{\tau}\left(R(v,e)-R^*\right), \\ \dfrac{\partial \rho C_v^*}{\partial t} + \dfrac{\partial \rho C_v^* u}{\partial x} = \dfrac{\rho}{\tau}\left(C_v(v,e)-C_v^*\right), \\ \dfrac{\partial \rho q^*}{\partial t} + \dfrac{\partial \rho q^* u}{\partial x} = \dfrac{\rho}{\tau}\left(q(v,e)-q^*\right), \\ \text{with } \tau \to 0^+. \end{cases} \quad (18)$$

Then, the source terms present on the right of the added equations (18) are considered through an instantaneous relaxation step, i.e., $\tau \to 0^+$. The functions $b(v,e)$, $P_\infty(v,e)$, $R(v,e)$, $C_v(v,e)$, and $q(v,e)$ are constructed in such a way that after the instantaneous relaxation step, the target EOS (TEOS) and PEOS are merged. Here, the target EOS is the BKW EOS. After the relaxation step, the extended equation model (14)-(18) thermodynamically closed with the predictor EOS NASG fundamentally corresponds to System (14) closed by the target EOS BKW. It remains to determine the various functions used during the relaxation step. Following the same methodology as in [10], the NASG parameters must fulfil the following coupling conditions between PEOS and TEOS,

$$\begin{cases} P_{TEOS}(v,e,y_i) = P_{PEOS}(v,e), \\ T_{TEOS}(v,e,y_i) = T_{PEOS}(v,e), \\ c^2_{TEOS}(v,e,y_i) = c^2_{PEOS}(v,e), \\ \Gamma_{TEOS}(v,e,y_i) = \Gamma_{PEOS}(v,e). \end{cases} \qquad (19)$$

The specific volume $v$ and internal energy $e$ are provided by the flow model (14). $\Gamma$ is the Grüneisen coefficient defined as $\Gamma = v \left.\frac{\partial P}{\partial e}\right|_v$. For the BKW formulation, the Grüneisen coefficient is,

$$\Gamma_{TEOS}(v,e,y_i) = v \left.\frac{\partial P_{TEOS}}{\partial e}\right|_{v,y_i} = \frac{R}{C_{v,TEOS}} \left( 1 + \left( 1 - \frac{\alpha T_{TEOS}}{T_{TEOS} + \theta}(1+\beta X) \right) X e^{\beta X} \right), \qquad (20)$$

with $C_{v,TEOS}$ given by equation (12). Comparing the formulation $P(v,T)$ for BKW (1) and NASG (17), the covolume $b^*$ is set to zero, since covolumes for the BKW formulation appears for each species in function $X$. System (19) is thus a set of 4 nonlinear equations with 4 unknown: $R^*$, $C_v^*$, $P_\infty^*$ and $q^*$. The solution to this system reads,

$$\begin{cases} C_v^* = \dfrac{R^*}{\Gamma_{TEOS}(v,e,y_i)}, \\[4pt] R^* = \dfrac{c^2_{TEOS}(v,e,y_i)}{(\Gamma_{TEOS}(v,e,y_i)+1)T_{TEOS}(v,e,y_i)}, \\[4pt] P_\infty^* = \dfrac{c^2_{TEOS}(v,e,y_i)}{v(\Gamma_{TEOS}(v,e,y_i)+1)} - P_{TEOS}(v,e,y_i), \\[4pt] q^* = e - \dfrac{c^2_{TEOS}(v,e,y_i)}{\Gamma_{TEOS}(v,e,y_i)} + v P_{TEOS}(v,e,y_i). \end{cases} \qquad (21)$$

Relations (21) are the functions used to reset the thermodynamic parameters during the relaxation step. These relations ensure that the coupling conditions (19) are fulfilled at any time and everywhere in the computational domain.

A major difference between the method of [10] and the present method is that the relaxation switch-off when convexity criteria are not fulfilled is absent with BKW. Indeed, during the numerous computations and tests done in this work with the BKW EOS, no numerical difficulties linked to convexity issues appeared (negative square sound speed for instance). The BKW EOS validity range seems very large, the EOS is quite robust and the relaxation switch-off is meaningless. Details on the convexity of the BKW EOS are given in Appendix A.

To validate the adaptation of the thermodynamic relaxation method to the BKW EOS, the numerical test cases considered until now are rerun with the present method. The results obtained previously with the conventional method (TEOS) and the relaxation method (PEOS) are compared and must be identical. Only the results for the NG shock-tube test of Figure 2 are shown in Figure 8 since conclusions are the same for all tests.

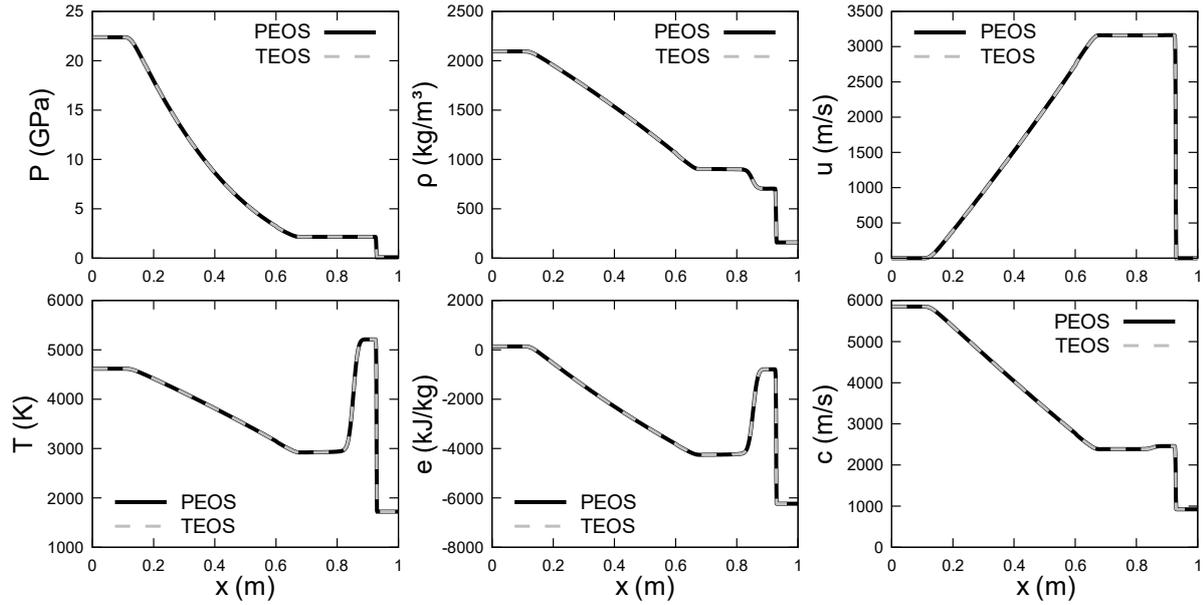

Figure 8. Shock tube test for the explosive NG of Figure 2 with the BKW EOS. Dashed grey lines represent computed results with the conventional method (TEOS), and black lines with the relaxation method (PEOS). The results of both methods are merged. The present thermodynamic relaxation method perfectly matches the target EOS results.

Figure 8 shows that results obtained previously with the conventional method, and with the present thermodynamical relaxation method are in perfect agreement. The method extended to the BKW EOS seems to work nicely. The thermodynamic relaxation method present no advantage regarding computational time when used in the flow model (14). Computations are slightly slower because of the added transport equations (18). However, in the next Sections 4 and 5, the treatment of solid carbon, which is governed by its own EOS, is addressed. The addition of condensed species implies that the flow model (14) is no longer sufficient. More complex multiphase flow models are needed. These models require numerical ingredients, such as pressure relaxation procedures, which are computationally costly. As is, it will be too expensive to use the BKW EOS for multiphase flows. The thermodynamic relaxation method goal is to lower the related cost while simplifying its implementation in numerical codes. Now that the method appears to work nicely for the BKW EOS, the treatment of solid carbon is addressed.

## 4. Cochran-Chan equation of state for solid carbon

All explosives considered above do not produce condensed carbon during their combustion as their oxygen balance is positive. However, most explosives have a negative oxygen balance (TNT, RDX, PETN, …). Associated detonation products contain solid carbon. Since the BKW EOS is appropriate only for gas mixtures, condensed species are modelled individually with their own equation of state. A mixture model is then considered with the BKW EOS for the gas mixture and other EOSs for condensed species.

The literature provides many EOSs to represent condensed species, especially carbon, in the detonation products. Cowan and Fickett [18] developed a sophisticated formulation, used for instance by Mader [14]. It reads,

$$P(v,T) = P_1(v) + a(v)T + b(v)T^2,$$
with $P_1(v) = -2.467 + 6.769\eta - 6.956\eta^2 + 3.040\eta^3 - 0.3869\eta^4,$
$a(v) = -0.2267 + 0.2712\eta,$
$b(v) = 0.08316 - 0.07804\eta^{-1} + 0.03068\eta^{-2}.$ (22)

$P$ is expressed in megabars and $T$ is in volt, i.e., in units of 11605.6 K. The term $\eta = v_0/v$ is the specific volume ratio with $v_0$ the specific volume of graphite in atmospheric conditions. A somewhat similar form is present in the Cheetah code [27] used in the present work,

$$v(P,T) = (4.99259 + 3.9828 \times 10^{-5}T + 1.191359 \times 10^{-9}T^2)$$
$$+ P(-6.377527 \times 10^{-6} + 1.1924995 \times 10^{-10}T - 3.7557816 \times 10^{-15}T^2)$$
$$+ P^2(3.58287 \times 10^{-12} - 1.00976 \times 10^{-16}T). \quad (23)$$

A modified Murnaghan equation of state was later implemented in Cheetah [32] to model condensed species for non-ideal explosives,

$$v(P,T) = v_0 \left( n\kappa P + e^{-\alpha(T-T_0)} \right)^{-1/n}. \quad (24)$$

The constants $n$, $\kappa$ and $\alpha$ are empirically fitted parameters and $T_0$ is the temperature in the reference state. Three sets of coefficients depending on the condensed carbon state (graphite, diamond, or liquid) are used. Adjustment of these EOS parameters is mostly based on experimental data points along the Hugoniot curve. Recent works aim to model condensed carbon under other states with phase transition. More details on available methods are given in [33]. Another important remark is that the condensed carbon EOS adjustment varies depending on the BKW parameters set chosen. For example, [22] use different graphite EOS for BKWC, BKWS and BKWR set of parameters. Last, in these references, only the thermal EOS is presented and no details on the associated caloric EOS are given.

The equations of state present in the literature, usually used in thermochemical codes, are quite sophisticated and ill adapted for intensive use in flow solvers. In the present contribution, the condensed carbon is considered as graphite and governed by the Cochran and Chan [11] (CC) equation of state under Mie-Grüneisen (MG) form. This EOS has been developed to model the solid reactant in the detonation reaction zone, consequently it should be able to model solid carbon to some extent. The CC EOS formulation and parameters adjustment are presented hereafter in Sections 4.1 and 4.2.

*4.1. Equation of state formulations*

The CC EOS [11] is expressed under MG form. The various formulations are summarized below,

$$\begin{cases} P(v,e) = \dfrac{\Gamma}{v}(e - e_k(v)) + P_k(v), \\ P(v,T) = \dfrac{\Gamma C_v T}{v} + P_k(v), \\ e(v,T) = C_v T + e_k(v), \\ c^2(v,P) = v(\Gamma + 1)(P - P_k(v)) - v^2 \dfrac{dP_k}{dv}. \end{cases} \quad (25)$$

The parameters $\Gamma$ and $C_v$ represent respectively the Grüneisen coefficient and the specific heat at constant volume. Functions $P_k$ and $e_k$ are specific to the MG-type EOS under consideration. For the CC EOS, these functions read,

$$\begin{cases} P_k(v) = A_1 \left(\dfrac{v}{v_{ref}}\right)^{-E_1} - A_2 \left(\dfrac{v}{v_{ref}}\right)^{-E_2}, \\ e_k(v) = -\dfrac{A_1 v_{ref}}{(1-E_1)}\left(\left(\dfrac{v}{v_{ref}}\right)^{1-E_1} - 1\right) + \dfrac{A_2 v_{ref}}{(1-E_2)}\left(\left(\dfrac{v}{v_{ref}}\right)^{1-E_2} - 1\right) - C_v T_{ref} + e_{ref}, \\ \dfrac{dP_k}{dv} = -\dfrac{A_1 E_1}{v_{ref}}\left(\dfrac{v}{v_{ref}}\right)^{-1-E_1} + \dfrac{A_2 E_2}{v_{ref}}\left(\dfrac{v}{v_{ref}}\right)^{-1-E_2}, \end{cases} \quad (26)$$

where $A_1$, $A_2$, $E_1$, and $E_2$ are empirically adjusted parameters. Subscript "ref" indicates the reference state data used for calibration. Adjustment of these parameters is addressed below.

### 4.2. Thermodynamic parameters

The CC EOS parameters are adjusted to satisfy the shock Hugoniot experimental data of the considered material. For solid carbon under graphite form, experimental values are obtained from online database [34].

Following the method of [11], parameters $\Gamma$, $A_1$, $A_2$, $E_1$, and $E_2$ are fitted along the linear curve $D = c_0 + su$, where $D$ is the shock speed, $u$ is the material velocity behind the shock, $c_0$ is the sound speed of the material at atmospheric conditions, and $s$ is the slope of the Hugoniot.

The reference energy $e_{ref}$ is adjusted so that the phase internal energy in standard state $e(P_{ref}, T_{ref})$, i.e., atmospheric conditions, is zero, since the enthalpy of formation of solid carbon is zero. The specific heat at constant volume $C_v$ is empirically calibrated based on CJ computations for TNT. More details are given in Section 5.1. The adjusted parameters for condensed carbon are summarized in Table 5 and fitting results are shown in Figure 9.

| Parameter | Value |
|---|---|
| $\Gamma$ | 0.277469 |
| $C_v$ (J/kg/K) | 2000 |
| $A_1$ (GPa) | 24.6643 |
| $A_2$ (GPa) | 25.0365 |
| $E_1$ | 2.5011 |
| $E_2$ | 1.5011 |
| $\rho_{ref}$ (kg/m$^3$) | 2250 |
| $T_{ref}$ (K) | 298.15 |
| $e_{ref}$ (J/kg) | -42.89 |

Table 5. CC EOS thermodynamic parameters for graphite, adjusted with experimental data from Rusbank database [34].

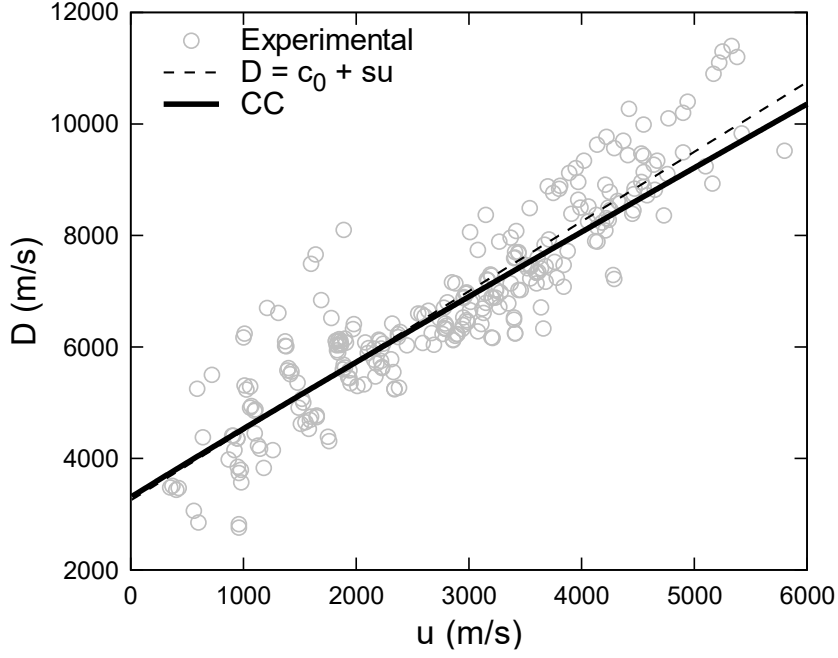

Figure 9. Shock wave speed versus material velocity for graphite. Experimental data from Rusbank are indicated with grey circles. The linear function $D = c_0 + su$ used for fitting the CC EOS parameters is represented with the dashed black line. Computed results with the CC EOS and parameters of Table 5 are indicated with the solid black line. The agreement is correct.

Having in hands the equation of state for solid carbon, computation of detonation products involving condensed carbon, i.e., negative oxygen balance explosives, is possible and investigated in the next Section 5.

## 5. BKW with condensed carbon

In this section, the modelling of detonation products involving condensed carbon is investigated. Detonation products are described through two distinct EOSs. The BKW EOS (Section 2) for the gas mixture and the CC EOS (Section 4) for solid carbon. The two-phase mixture BKW-CC is considered in both pressure and temperature equilibrium, as done in all thermochemical codes. It means that the mixture specific volume and mixture internal energy are given by,

$$v = Y_S v_S(P,T) + Y_G v_G(P,T,y_i), \tag{27}$$

$$e = Y_S e_S(P,T) + Y_G e_G(P,T,y_i). \tag{28}$$

S denotes the condensed phase governed by CC, and G denotes the gas mixture governed by BKW. $Y$ corresponds to the mass fraction in the detonation products mixture (G+S) ($Y_S + Y_G = 1$). Each gas species $i$ present in BKW also has its own mass fraction $Y_i$ in the detonation products mixture so that $Y_G = \sum Y_i$. Computations with the BKW formulations requires the species mass fractions $y_i$ inside the gas part of the detonation products. Both mass fractions are linked by the relation,

$$y_i = \frac{Y_i}{Y_G}. \tag{29}$$

The accuracy of this thermodynamic model is assessed with various test cases. In Section 5.1, CJ states are computed and compared to both experimental data and Cheetah computations for various explosives. Predicted results appear accurate. In Section 5.2, 1D shock-tube tests between detonation products and air are achieved and compared to results obtained with the JWL EOS. The multiphase model of [13] detailed in Appendix C is used both to solve the mixture model BKW-CC (27)-(28), and treat the interface between JWL and air. Predictions with BKW-CC and JWL appear in good agreement. Mass transfer between the solid carbon phase and gas phase is modelled in Section 5.3 to account for variations in the detonation products composition. In Section 5.4, spherical explosions of TNT in air are studied and predictions are compared to experimental data of pressure signals. Lastly, in Section 5.5, a simple post-combustion model is developed to show the BKW-CC thermodynamic model capabilities.

*5.1. CJ state for various explosives*

The CJ state for various explosives is computed in this section. The method used to compute CJ state with the mixture model (27)-(28) is summarized in Appendix D. Results are compared to experimental data from [8], except for AN where experimental data are from [35]. Computed results are also compared with the thermochemical code Cheetah 2.0 results [36] using the BKWC library. The initial density and reference energy for each reactive material, needed to compute the CJ state, are summarized in Table 6. The composition at the CJ state computed with Cheetah and used in BKW-CC computations are also given.

| Explosive | $e_0$ kJ/kg | $\rho_0$ kg/m$^3$ | $CO_2$ | $H_2O$ | $N_2$ | $O_2$ | $CO$ | $NO$ | $C_{(s)}$ |
|---|---|---|---|---|---|---|---|---|---|
| AN | -4562 | 1050 | 0 | 0.450 | 0.350 | 0.200 | 0 | 0 | 0 |
| HMX | 252 | 1890 | 0.281 | 0.243 | 0.379 | 0 | 0.020 | 0 | 0.077 |
| HMX | 252 | 1600 | 0.237 | 0.243 | 0.379 | 0 | 0.077 | 0 | 0.064 |
| NG | -1640 | 1600 | 0.565 | 0.198 | 0.180 | 0.036 | 0.010 | 0.010 | 0 |
| NM | -1753 | 1130 | 0.087 | 0.418 | 0.233 | 0 | 0.171 | 0 | 0.091 |
| PETN | -1667 | 1760 | 0.500 | 0.228 | 0.177 | 0 | 0.072 | 0 | 0.022 |
| PETN | -1667 | 1600 | 0.457 | 0.227 | 0.177 | 0 | 0.128 | 0 | 0.010 |
| RDX | 311 | 1800 | 0.271 | 0.243 | 0.378 | 0 | 0.033 | 0 | 0.074 |
| RDX | 311 | 1400 | 0.190 | 0.248 | 0.391 | 0 | 0.121 | 0 | 0.049 |
| TATB | -562 | 1850 | 0.243 | 0.209 | 0.326 | 0 | 0.016 | 0 | 0.206 |
| TNM | 188 | 1640 | 0.210 | 0 | 0.264 | 0.455 | 0 | 0.072 | 0 |
| TNT | -277 | 1640 | 0.274 | 0.198 | 0.185 | 0 | 0.084 | 0 | 0.259 |
| TNT | -277 | 1450 | 0.226 | 0.196 | 0.185 | 0 | 0.147 | 0 | 0.245 |
| TNT | -277 | 1360 | 0.201 | 0.194 | 0.186 | 0 | 0.182 | 0 | 0.237 |
| TNT | -277 | 1000 | 0.111 | 0.169 | 0.187 | 0 | 0.345 | 0 | 0.188 |
| TNT | -277 | 800 | 0.078 | 0.134 | 0.189 | 0 | 0.452 | 0 | 0.147 |

Table 6. Energy and initial density for each explosive material considered. The composition at the CJ state computed with Cheetah using the BKWC library is given and used for the computations with the BKW-CC mixture model.

The computed CJ state (detonation velocity, pressure, and temperature) is summarized in Table 7.

| Explosive | $\rho_0$ kg/m$^3$ | Experimental | | | Cheetah | | | BKW-CC | | |
|---|---|---|---|---|---|---|---|---|---|---|
| | | D (m/s) | P (GPa) | T (K) | D (m/s) | P (GPa) | T (K) | D (m/s) | P (GPa) | T (K) |
| AN | 1050 | 4500 | | | 4329 -3.8% | 4.66 | 1688 | 4241 -5.8% | 4.43 | 1619 |
| HMX | 1890 | 9110 | 39.0 | | 9244 1.5% | 38.6 -1.0% | 4124 | 9403 3.2% | 36.62 -6.1% | 4121 |
| | 1600 | 7910 | 28.0 | 4300 | 8162 3.2% | 26.2 -6.4% | 4281 -0.4% | 8192 3.6% | 24.79 -11.5% | 4271 -0.7% |
| NG | 1600 | 7700 | 25.3 | 4260 | 7701 0.0% | 22.4 -11.5% | 4518 6.1% | 7666 -0.4% | 22.30 -11.9% | 4553 6.9% |
| NM | 1130 | 6350 | 12.5 | 3430 | 6064 -4.5% | 11.4 -8.8% | 3664 6.8% | 6031 -5.0% | 10.48 -16.2% | 3760 9.6% |
| PETN | 1760 | 8270 | 33.7 | | 8513 2.9% | 30.8 -8.6% | 4360 | 8497 2.7% | 28.92 -14.2% | 4342 |
| | 1600 | 7750 | 26.6 | 4400 | 7973 2.9% | 25.3 -4.9% | 4410 0.2% | 7934 2.4% | 23.55 -11.5% | 4384 -0.4% |
| RDX | 1800 | 8750 | 34.1 | | 8920 1.9% | 34.5 1.2% | 4213 | 9025 3.1% | 32.69 -4.1% | 4206 |
| | 1400 | 7460 | 21.3 | 4610 | 7479 0.3% | 19.9 -6.6% | 4361 -5.4% | 7465 0.1% | 18.75 -12.0% | 4367 -5.3% |
| TATB | 1850 | 7660 | 25.9 | | 7824 2.1% | 27.2 5.0% | 3288 | 8019 4.7% | 25.44 -1.8% | 3242 |
| TNM | 1640 | 6360 | 15.9 | 2800 | 6267 -1.5% | 13.8 -13.2% | 2652 -5.3% | 6093 -4.2% | 12.64 -20.5% | 2371 -15.3% |
| TNT | 1640 | 6930 | 21.0 | | 6843 -1.3% | 19.2 -8.6% | 3712 | 6967 0.5% | 18.11 -13.8% | 3676 |
| | 1450 | 6500 | 14.4 | | 6266 -3.6% | 14.3 -0.7% | 3698 | 6345 -2.4% | 13.68 -5.0% | 3675 |
| | 1360 | 6200 | 12.4 | | 5995 -3.6% | 12.5 -0.7% | 3681 | 6061 -2.2% | 11.89 -4.1% | 3666 |
| | 1000 | 5000 | 6.7 | 3400 | 4983 -0.3% | 6.9 3.0% | 3557 4.6% | 4973 -0.5% | 6.37 -4.9% | 3643 7.1% |
| | 800 | 4340 | 3.7 | | 4475 3.1% | 4.7 27.0% | 3432 | 4404 1.5% | 4.24 14.6% | 3658 |

Table 7. Detonation velocity, pressure, and temperature at the CJ point for various explosives at different initial densities. Relative errors to experimental data are also given. Experimental data are taken from [8,35].

As stated in Section 4.2, the specific heat $C_v$ for the CC EOS of solid carbon is empirically adjusted based on CJ computations for TNT. TNT is chosen as its detonation products composition involves the highest amount of solid carbon of all considered explosives. CJ computations for TNT with the initial density of $1640 \text{ kg/m}^3$ are considered for various values of $C_v$ and the most accurate one with respect to computed detonation velocity is retained. After multiple trials, the retained specific heat is $C_v = 2000$ J/kg/K, and is employed for all explosives. This empirical adjustment of $C_v$ seemed accurate enough to model solid carbon in the various test cases considered in this work. It is worth noting that computed temperatures are closed to experimental ones reported by [8].

CJ states computed with the mixture model BKW-CC are in correct agreement with experimental data for the various explosives and initial density considered in Table 7. Relative errors are comparable to those of Cheetah using the BKWC library. These results seem accurate enough to validate the ideal gas energy contribution (8) and the solid carbon CC EOS (Section 4) used and calibrated herein.

*5.2. Shock-tube tests*

The mixture model BKW-CC is tested in this Section in the context of shock-tube computations similar to those studied in Section 3.4. A multiphase flow is needed since the detonation products are modelled with two distinct phases: gas products, governed by the BKW EOS and solid carbon governed by the CC EOS. The multiphase model considered in this frame is the one of Saurel et al. [13] as it deals with mixtures in velocity and pressure equilibrium, while temperature relaxation can be modelled with stiff or finite rate relaxation. This model is also able to deal with interface computations, as done previously with the JWL EOS of Figure 7.

Special care is needed to initialize both phases at equilibrium for a given initial pressure $P$, detonation product density $\rho$ and detonation products composition $Y_k$ for $k \in \{G, S\}$. The individual density of each phase BKW (G) and CC (S) as well as volume fractions needs to be determined. The mixture equilibrium temperature is determined with an iterative method using the mass conservation of the mixture (27) by solving the function,

$$f(T) = Y_S v_S(P,T) + Y_G v_G(P,T,y_i) - v. \qquad (30)$$

With the solution temperature, individual specific volumes are determined with $v_k = v_k(P,T)$, and volume fraction are computed as follows,

$$\alpha_k = Y_k \frac{v_k}{v}. \qquad (31)$$

The first tests consider 1D shock-tube tests for various negative oxygen balance explosives given in Table 7. The high-pressure chamber initially consists of detonation products at CJ conditions, and the low-pressure chamber contains air at atmospheric conditions. With the BKW-CC model, the air is modelled with the BKW EOS by setting the air composition $Y_{N_2} = 0.74$ and $Y_{O_2} = 0.26$. With the JWL EOS, the air is governed by the ideal gas EOS with $\gamma = 1.4$ and $C_v = 719$ J/kg/K.

Simulations are first-order accurate in both space and time. The CFL criterion is 0.8. The computational domain is 1-m long and spatially resolved with 1000 cells. All phases are in velocity and pressure equilibrium. Additionally, the detonation products (BKW-CC) are in thermal equilibrium through a stiff temperature relaxation solver [37]. Both solid and gas phases are also treated with the thermodynamic relaxation method of [10], detailed for the CC EOS in the same reference and in Section 3.5 for the BKW EOS.

The computed results for TNT, RDX and PETN are shown in Figure 10 with each column corresponding to each explosive respectively. Each row corresponds to computed flow results: pressure, velocity, density and temperature.

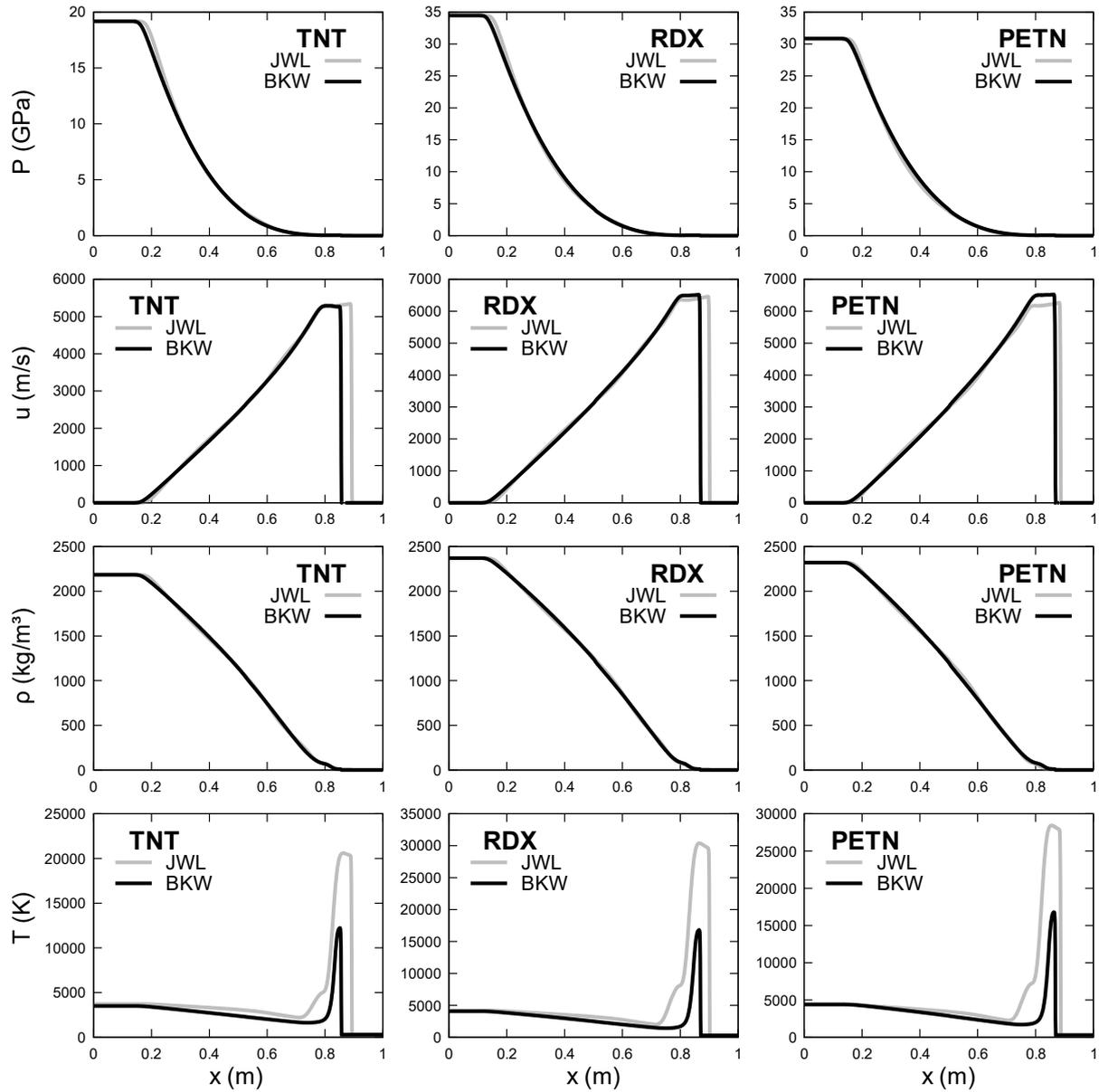

Figure 10. Shock-tube test for the TNT, RDX and PETN detonation products in contact with air. The left state corresponds to the detonation products at CJ conditions, and the right state corresponds to air at atmospheric conditions. The initial discontinuity is located at 0.5 m. Results are shown at time 60 µs, 50 µs and 50 µs respectively. Grey lines represent computed results with the JWL EOS, and black lines with the BKW-CC EOSs.

Computed results with the JWL EOS and the BKW-CC EOSs for the detonation products are close for all explosives considered. The main differences are on the temperature and the shock position. These differences are caused by the treatment of the air interface which is different with the JWL and BKW-CC. Indeed, with the JWL EOS, air is governed by the ideal gas EOS, while air is governed by the BKW EOS in the other case. Moreover, with the JWL EOS, air is considered as a separate phase only in mechanical equilibrium with detonation products. While in the BKW-CC case, air being modelled by the same BKW, it is inherently in both mechanical and thermal equilibrium with detonation products. In order to strictly compare the behaviour of detonation products with the JWL EOS and BKW-CC model, the BKW-CC shock-tube tests are rerun with air modelled by the ideal gas EOS as an additional separate phase. Computed results are shown in Figure 11.

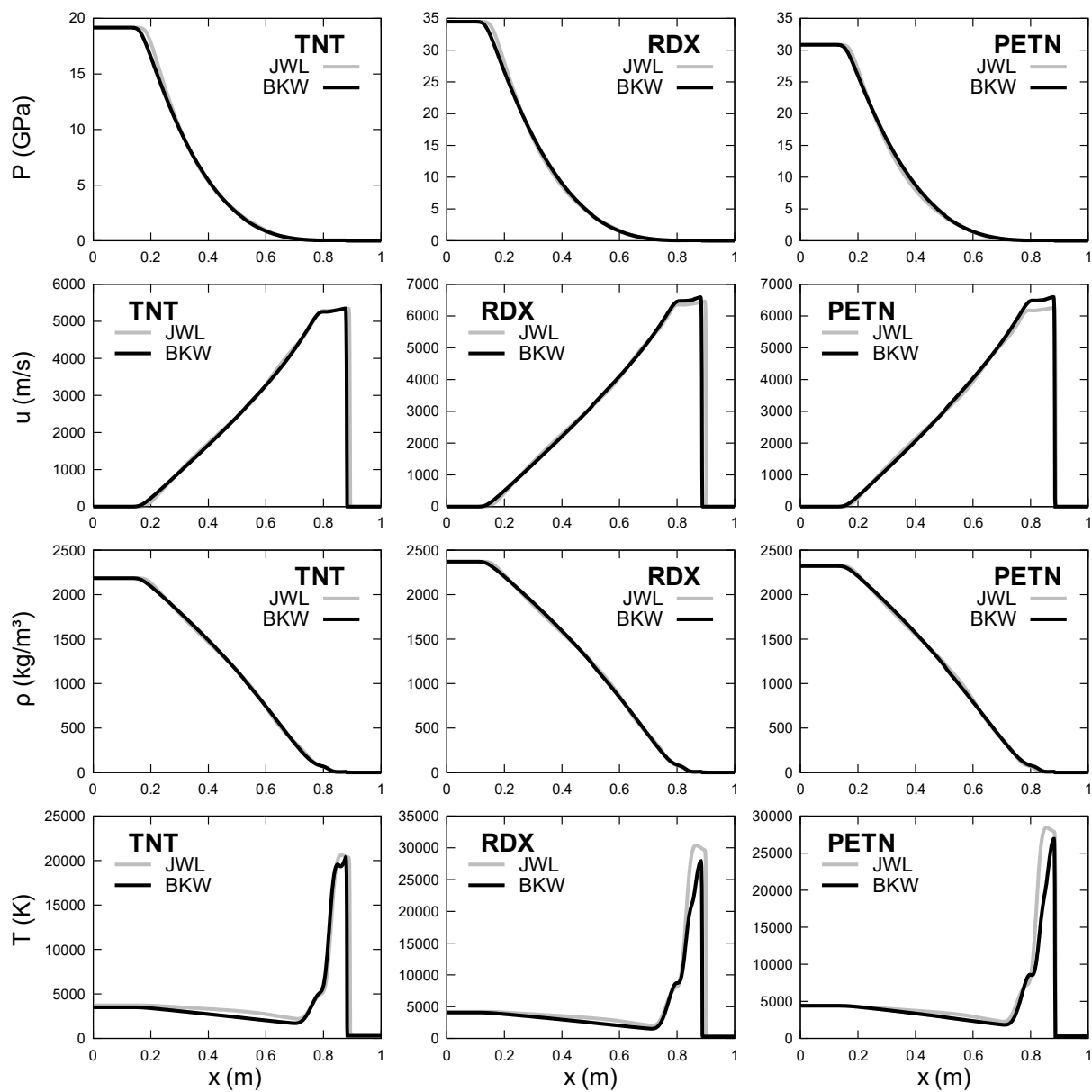

Figure 11. Shock-tube tests of Figure 10 with air modelled as a separate phase governed by the ideal gas EOS for both JWL and BKW-CC EOSs. Grey lines represent computed results with the JWL EOS, and black lines with the BKW-CC EOSs.

With the same air thermodynamics, JWL and BKW-CC EOSs predictions around the interface are much closer. Indeed, in Figure 11 the shock position and temperatures close to the interface are in better agreement compared to previous results in Figure 10. Differences similar to the observations of Section 3.2 for shock-tube tests with positive oxygen balance explosives are observed. Slight discrepancies mainly on velocity plateau and temperatures are present but remain acceptable. These results validate the detonation products representations with two-phases, BKW for gas products and CC for solid carbon, in both mechanical and thermal equilibrium. Indeed, in the present computations, the JWL EOS implicitly contains solid carbon in the same mechanical and thermal equilibrium conditions.

In these tests, the detonation products composition is constant. This is obviously not the case as already discussed in Section 3.3. The addition of condensed carbon complicates the model developed earlier, valid only for gas species variations. An extra mass transfer model is developed in the following section.

### *5.3. Mass transfer*

As studied in Section 3.3, the detonation products composition depends on thermodynamic conditions. To account for these variations, the same methodology is considered. However, since solid carbon is a separate phase, an appropriate mass transfer model between the BKW and CC phases is needed. Following the mass transfer modelling derived in [38,39], the volume fraction and mass equations for the solid carbon (S) and gas mixture (G) are modified with additional source terms,

$$\begin{cases} \dfrac{\partial \alpha_G}{\partial t} = -\dfrac{\rho \dot{Y}_S}{\rho_I}, \\[4pt] \dfrac{\partial \alpha_G \rho_G}{\partial t} = -\rho \dot{Y}_S, \\[4pt] \dfrac{\partial \rho Y_i}{\partial t} = \rho \dot{Y}_i, \quad \forall i, \\[4pt] \dfrac{\partial \alpha_S}{\partial t} = \dfrac{\rho \dot{Y}_S}{\rho_I}, \\[4pt] \dfrac{\partial \alpha_S \rho_S}{\partial t} = \rho \dot{Y}_S. \end{cases} \qquad (32)$$

Most of the notations are detailed in Appendix C. The mixture density is $\rho = \alpha_G \rho_G + \alpha_S \rho_S$. $\rho_I$ is the interfacial density between the solid and gas phase approximated as in [38],

$$\rho_I = \dfrac{\dfrac{\rho_S c_S^2}{\alpha_S} + \dfrac{\rho_G c_G^2}{\alpha_G}}{\dfrac{c_S^2}{\alpha_S} + \dfrac{c_G^2}{\alpha_G}}. \qquad (33)$$

The source terms $\dot{Y}$ are defined for each gas species $i$ and solid carbon as,

$$\dot{Y}_i = \dfrac{Y_i^*(\rho) - Y_i}{\Delta t}. \qquad (34)$$

$\Delta t$ indicates the time step during which the mass transfer is evolved. $Y_i^*(\rho)$ is the density dependent function (16) which estimates the detonation product composition linearly between two points. This function is recalled hereafter,

$$Y_i^*(\rho) = Y_{i,1} + (Y_{i,2} - Y_{i,1}) \dfrac{\rho - \rho_1}{\rho_2 - \rho_1}. \qquad (35)$$

When air is present and modelled with the BKW EOS, the source terms need to be modified. Indeed, the compositions at the two different points $Y_{i,1}$ and $Y_{i,2}$ computed with Cheetah does not account for air in the detonation products composition. In this work, the air is assumed inert during the mass transfer update. To make distinction between the species $N_2$ and $O_2$ present in the detonation products, considered by mass transfer effects, and the same species which constitutes air, two new species $N_{2,in}$ and $O_{2,in}$ are introduced. These species are inert (in) during the mass transfer, i.e., $\dot{Y}_{N_{2,in}} = \dot{Y}_{O_{2,in}} = 0$, and are used to model air with the BKW EOS. The thermodynamic parameters for the inert species are identical to their counterpart. At the interface at least, because of numerical diffusion, inert air coexists with detonation products. As stated, the target composition $Y_i^*$ of the mass transfer does not contain air. To make sure that the saturation constraint $\sum Y_i = 1$ is still satisfied after the mass transfer when air is present, the target composition is adjusted as follows,

$$Y_i^*(\rho) = \left( Y_{i,1} + (Y_{i,2} - Y_{i,1}) \frac{\rho - \rho_1}{\rho_2 - \rho_1} \right) \left( 1 - Y_{N_{2,in}} - Y_{O_{2,in}} \right). \tag{36}$$

The mass transfer step is achieved last, after the hyperbolic step, after the thermodynamic relaxation, and after both temperature and pressure relaxation. The mass transfer source terms (32) changes instantaneously the detonation products composition according to the function (35), or (36) when air is present in BKW, while guaranteeing mass conservation of detonation products.

The effect of mass transfer on computed results is evaluated over the shock-tube test presented in Figure 10 for TNT. The air is modelled in the BKW EOS with the BKW-CC model, and as a separate ideal gas phase with JWL EOS. The mass fractions for each species and densities in CJ conditions and atmospheric conditions needed for the function (36) are summarized in Table 8.

| | Parameter | CJ | Atmospheric |
|---|---|---|---|
| | $\rho\,(\text{kg/m}^3)$ | 2185 | 1.35 |
| | $CO_2$ | 0.273855 | 0.190573 |
| | $H_2O$ | 0.197630 | 0.158593 |
| $Y_i$ | $N_2$ | 0.185212 | 0.187172 |
| | $CO$ | 0.083813 | 0.261319 |
| | $NO$ | 0.000068 | 0.000001 |
| | $C_{(s)}$ | 0.259420 | 0.202340 |

Table 8. Density and mass fractions of the detonation product species for TNT at the CJ point and atmospheric point. Data are computed with Cheetah using the BKWC library.

Results are given in Figure 12 for the JWL EOS and BKW-CC with and without mass transfer.

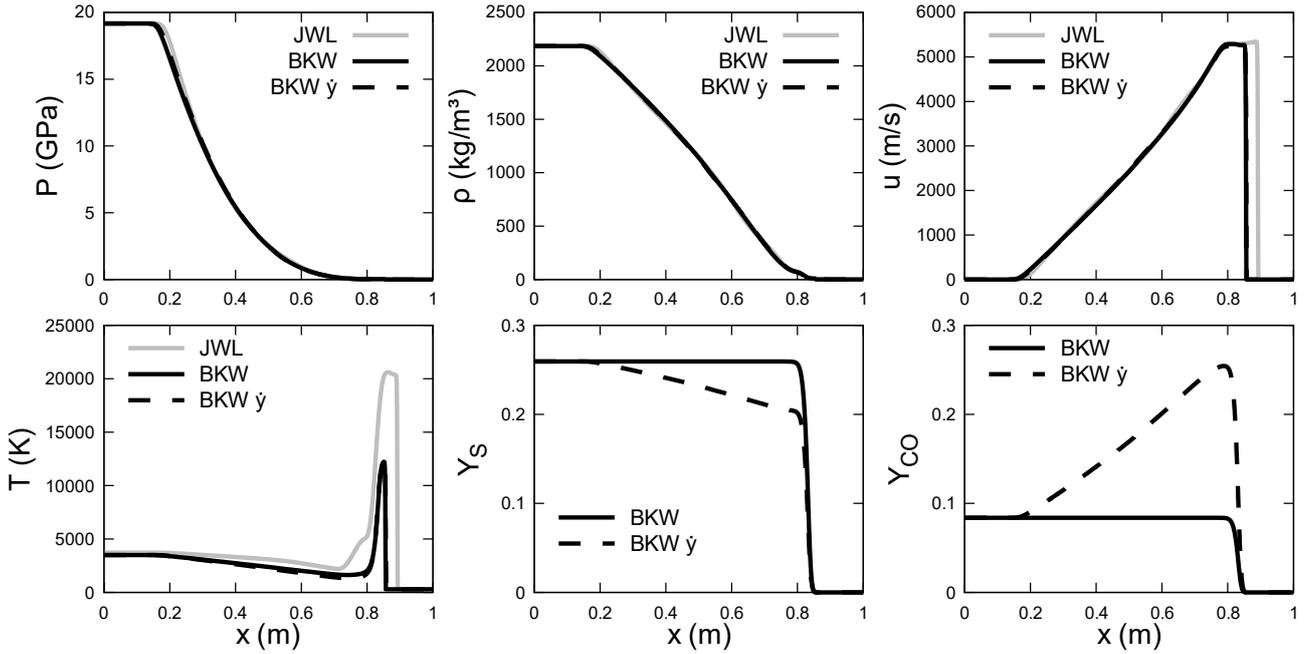

Figure 12. Shock-tube test for the explosive TNT in air. The initial discontinuity is located at 0.5 m. Results are shown at time 60 µs. Grey lines represent results with the JWL EOS. Black lines represent results with the BKW EOS with constant composition, and dashed lines with mass transfer. For both BKW computations, computed mixture variables are quite close.

From the mass fractions of the solid carbon and gas species $CO$ shown in Figure 12, the mass transfer model appears to work as intended. Mass fractions vary gradually from the CJ composition to the atmospheric one, following density variations. Nonetheless, the mass transfer has only a sensible impact on predicted results for pressure, velocity and temperature. As mentioned in Section 3.3 for positive oxygen balance explosives, the composition of detonation products does not vary much between the CJ point and atmospheric conditions. This appears to be the same for negative oxygen balance explosives. This fact is well known and reported as Kamlet and Jacobs rules [40].

The following test is a spherical TNT explosion in air. Mass transfer between phases is activated as with spherical effects, detonation products quickly reach atmospheric conditions, and the chemical composition varies faster.

### *5.4. Spherical explosion in air*

In this section, a spherical TNT explosion in air is studied. The experimental setup and related data are given in [30]. A spherical charge of 2.8 kg with radius 7.36 cm, is initialized at its center. Pressure gauges are located at 1 m, 2 m, and 3 m from the charge center. Numerical computations are addressed in 1D with spherical symmetry. The domain is 4-m long with 1 mm cells. Computations are first order in space and time with CFL number 0.8. The final time is 4 ms. The charge is initialized with a constant volume explosion state corresponding to $P = 8.67$ GPa and $\rho = 1654$ kg/m$^3$, computed with Cheetah. Pressure signals at the various gauges for the JWL and BKW-CC EOSs are compared to experimental records in Figure 13.

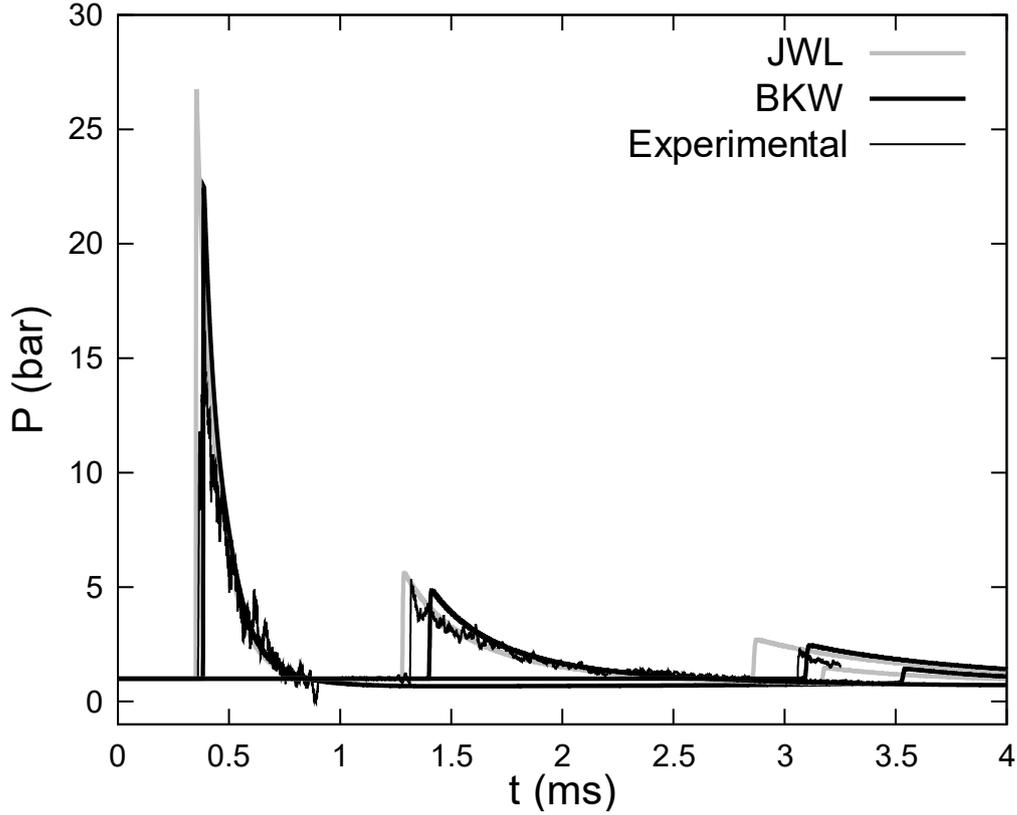

Figure 13. Pressure signals for a spherical TNT charge explosion in air. Sensors are placed at 1 m, 2 m, and 3 m from the charge center. Fine black lines represent experimental data. Grey lines represent computed results with the JWL EOS, and thick black lines with the BKW-CC EOSs and mass transfer.

Predictions are comparable between the JWL EOS and the BKW-CC model. Both exceed the experimental pressure signal at the first sensor, but BKW-CC is in better agreement. On the second and third sensors, computed pressure signals are quite satisfying with both thermodynamic models. The maximal pressure is slightly improved with BKW, particularly on the third sensor. This improvement may come from the wider range of validity of BKW compared to JWL. The arrival time is faster with JWL than BKW, which seems in better agreement.

CPU time of the BKW-CC formulation is compared to JWL for the present spherical air explosion test in Table 9.

|         | With thermodynamic relaxation | Without thermodynamic relaxation |
|---------|-------------------------------|----------------------------------|
| JWL     | 3 min 1 s                     | 6 min                            |
| BKW-CC  | 6 min 37 s                    | 1 h 1 min 32 s                   |

Table 9. Computational times for 1D spherical TNT explosions in air, presented in Figure 13. Computations are done with and without the thermodynamic relaxation method of [10].

The BKW-CC EOS with thermodynamic relaxation computation time is comparable to the conventional resolution with JWL EOS. While the JWL EOS with the relaxation method is accelerated by a factor 2, the BKW-CC model is accelerated by a factor 7, showing the importance of the thermodynamic relaxation method. This method renders BKW and its extensions, such as BKW-CC appropriate for multidimensional hydrocodes.

The next Section 5.5 illustrates the BKW-CC model capabilities to account for secondary reactions, such as post-combustion.

## 5.5. Post-combustion

In this section, a simplified model of post-combustion reaction is developed, mainly to show the BKW-CC ability to consider secondary reactions in detonation products. Post-combustion effects involve detonation products burning in air. Only the combustion of solid carbon is considered here,

$$C_{(s)} + O_2 \rightarrow CO_2. \qquad (37)$$

This reaction is exothermic with an enthalpy of reaction: $\Delta \hat{H}_R = -393.5 \text{ kJ/mol}$.

With the BKW-CC EOS and the flow model of Appendix C, the composition is known everywhere thanks to the mass fractions $Y_k$. Similar source terms as for the mass transfer (32) are used. Only the species $O_2$, $CO_2$, and solid phase, which is carbon $C_{(s)}$, are impacted by the post-combustion reaction (37). All the other species are inert. Thus, it remains to determine the composition variation $\dot{Y}_i$ for each species of (37). Let's denote by $n_i$ the initial number of moles of each species $i$ present, and $n_i^*$ the number of moles after the post-combustion reaction (37). The number of moles is linked to the mass fraction with,

$$n_i = \frac{\rho Y_i}{W_i}. \qquad (38)$$

Two instances are possible depending on the local oxygen balance,

$$\text{if } n_{O_2} - n_{C_{(s)}} < 0, \begin{cases} n_{C_{(s)}}^* = n_{C_{(s)}} - n_{O_2} \\ n_{O_2}^* = 0 \\ n_{CO_2}^* = n_{CO_2} + n_{O_2} \end{cases}, \text{ otherwise, } \begin{cases} n_{C_{(s)}}^* = 0 \\ n_{O_2}^* = n_{O_2} - n_{C_{(s)}} \\ n_{CO_2}^* = n_{CO_2} + n_{C_{(s)}} \end{cases}. \qquad (39)$$

From post-reaction mole numbers, the mass fraction variation for each species ($O_2$, $CO_2$, and solid phase) is obtained,

$$\dot{Y}_i = \frac{Y_i^* - Y_i}{\Delta t} = \frac{W_i}{\rho \Delta t}\left(n_i^* - n_i\right). \qquad (40)$$

The post-combustion model presented is evaluated in the context of a basic shock-tube test. The computational domain is 1-m long and spatially resolved with 1000 cells. The interface is located at 0.5 m. The high-pressure chamber is at 10 bars and low-pressure at 1 bar. Both chambers are initially at 300 K. In the left state, a mixture of 0.8 $N_2$ and 0.2 $C_{(s)}$ modelled with the BKW-CC model is present. The right state contains air governed by BKW. The results at 0.6 ms are shown in Figure 14 with and without the combustion model.

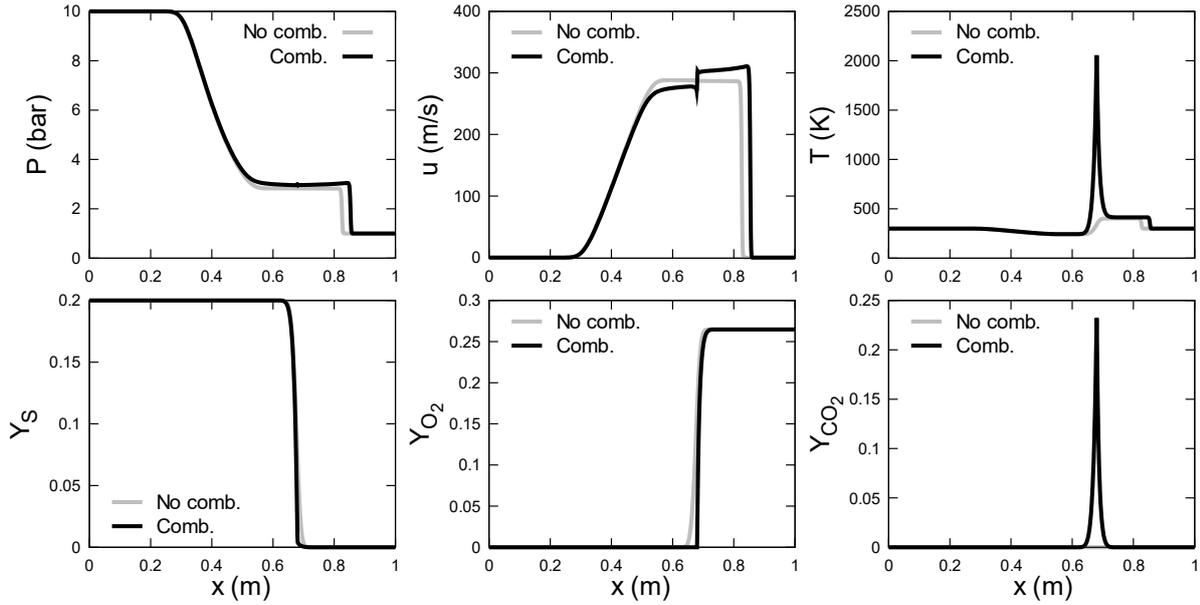

Figure 14. Shock-tube test illustrating the post-combustion model with the BKW-CC EOSs. A mixture of $N_2$ and $C_{(s)}$ in the left-state propagates in air. The combustion reaction (37) occurs at the interface and $CO_2$ is formed. The solution with and without the post-combustion model is represented respectively by the black and grey lines.

At the interface, mixing of $C_{(s)}$ and $O_2$ occurs, which allows the combustion reaction (37) to happen and form $CO_2$, as illustrated in Figure 14. The combustion reaction increases the temperature at the interface since the reaction is exothermic. It results in enhanced shock wave amplitude.

The presented post-combustion process is simplified in the aim to illustrate the capabilities with the BKW-CC EOSs.

## 6. Conclusion

In the present work, a thermodynamic model to represent detonation products with the BKW EOS for gas species and CC EOS for solid carbon has been developed. This formulation presents some advantages compared to the JWL EOS, which is the most popular EOS for detonation products.

- Most thermochemical codes use the BKW EOS to fit JWL parameters. The BKW formulation is explosive independent and only requires knowledge of the detonation products composition.
- The BKW EOS considers the gas species present in the detonation products. Mass fraction equations are present, and source terms have been modelled to account for composition variation. No sophisticated and computationally expensive chemical equilibrium solver is needed in the flow model.
- BKW has a range of validity and convexity larger than JWL, as it tends to the ideal gas EOS at low densities.
- The CC EOS is used for solid carbon and is a simpler formulation compared to the ones used in thermochemical codes.
- The multiphase model of Saurel et al. [13] is used to compute efficiently the detonation products BKW-CC in both mechanical and thermal equilibrium with a splitting method for stiff pressure and stiff temperature relaxations. This approach seemed more efficient than a flow model in strict temperature and pressure equilibrium.
- The thermodynamic relaxation method of Neron et al. [10] is adapted successfully to the BKW EOS to reduce CPU cost of the BKW-CC model, rendering computational costs comparable to JWL.

The BKW-CC model showed results similar to JWL while more flexible with access to the composition (interface with air, secondary reaction). Moreover, the BKW-CC model proved satisfying results for CJ computations, and better accuracy than JWL for temperature computations far from the CJ point. Moreover, the various computations done with the thermodynamic relaxation method showed significant robustness with BKW-CC.

## Appendix A. Convexity of the BKW EOS

The convexity of an equation of state requires fulfilment of various criteria [41,42],

$$\begin{cases} \left.\dfrac{\partial^2 e}{\partial v^2}\right|_s > 0, \\ \left.\dfrac{\partial^2 e}{\partial s^2}\right|_v > 0, \\ \left.\dfrac{\partial}{\partial s}\right|_v \left.\dfrac{\partial e}{\partial v}\right|_s > 0, \\ \left.\dfrac{\partial^2 e}{\partial s^2}\right|_v \left.\dfrac{\partial^2 e}{\partial v^2}\right|_s - \left(\left.\dfrac{\partial}{\partial s}\right|_v \left.\dfrac{\partial e}{\partial v}\right|_s\right)^2 > 0. \end{cases} \quad (A.1)$$

However, the formulation with entropy may be intricate to manipulate. This is especially the case for the BKW EOS for which the determination of the energy formulation $e(v,s)$ is quite challenging. To avoid such complexity, conditions (A.1) are expressed without entropy (see [43] Appendix A). With the help of the Gibbs identity and the Maxwell relations [20], the convexity conditions (A.1) are rearranged as:

$$\begin{cases} \dfrac{c^2}{v^2} > 0, \\ \left.\dfrac{\partial T}{\partial P}\right|_v \left(\left.\dfrac{\partial e}{\partial v}\right|_T + P\right)\left(\left.\dfrac{\partial e}{\partial T}\right|_v\right)^{-1} > 0, \\ \left(\left.\dfrac{\partial e}{\partial v}\right|_T + P\right)\left(\left.\dfrac{\partial e}{\partial T}\right|_v\right)^{-1} > 0, \\ \left(\left.\dfrac{\partial e}{\partial v}\right|_T + P\right)\left(\left.\dfrac{\partial e}{\partial T}\right|_v\right)^{-2}\left(\dfrac{c^2}{v^2}\left.\dfrac{\partial e}{\partial P}\right|_v - \left.\dfrac{\partial e}{\partial v}\right|_T - P\right) > 0. \end{cases} \quad (A.2)$$

The BKW formulations defined in Section **2.1** are inserted into (A.2). After some algebraic manipulations, the following sufficient condition appears,

$$\rho < \frac{(\theta + (1-\alpha)T)(T+\theta)^{\alpha}}{\alpha\beta\kappa T \sum \frac{y_i}{W_i} k_i}. \tag{A.3}$$

Condition (A.3) does not strictly correspond to the convexity criteria (A.2) for the BKW EOS, but is a sufficient condition to fulfil them.

The BKW parameters $\alpha$, $\beta$, $\theta$, $\kappa$, $k_i$, $a_i$, and $b_i$ are positive constants. Condition (A.3) is now analysed with the parameters given in Section **2.2**. It only depends on the temperature and the gas detonation products composition. For the sake of the analysis, the composition considered is the most restrictive, i.e., with the highest possible ratio $\sum y_i k_i / W_i$. This corresponds to a pure $N_2$ composition ($y_{N_2} = 1$), since its ratio $k_i / W_i$ is the highest with this species. Relation (A.3) is represented in Figure 15.

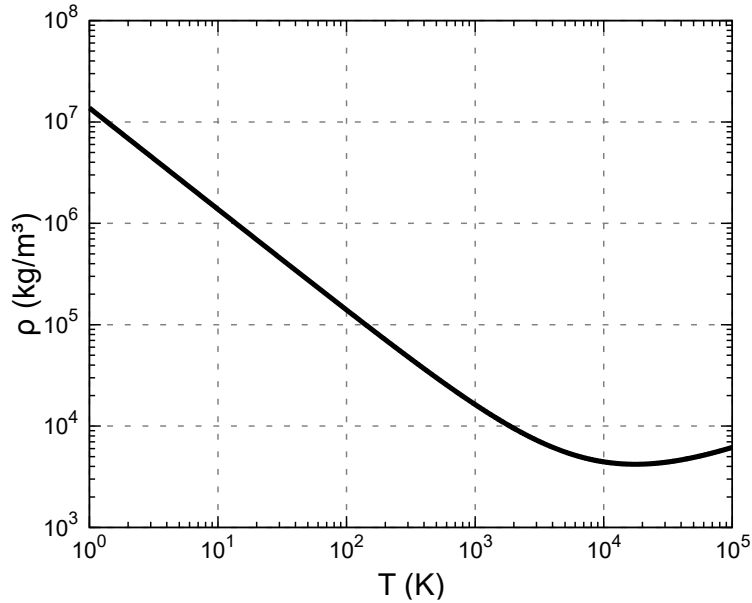

Figure 15. Sufficient convexity condition (A.3) in terms of temperature. For a given temperature, if the density is below the black curve, the BKW EOS is convex.

In Figure 15, the density needs to stay below the curve representing the condition (A.3) to ensure the convexity of the BKW EOS. From the high density and temperature values, it seems very unlikely to violate the convexity criteria with the BKW EOS. Indeed, the minimum density located at $17607$ K is $4202$ kg/m$^3$, which is much higher than the detonation products density of any explosive.

## Appendix B. JWL equation of state

The Jones-Wilkins-Lee (JWL) [1] equation of state, under Mie-Grüneisen (MG) form reads,

$$\begin{cases} P(v,e) = \dfrac{\Gamma}{v}\left(e - e_k(v)\right) + P_k(v), \\[6pt] P(v,T) = \dfrac{\Gamma C_v T}{v} + P_k(v), \\[6pt] e(v,T) = C_v T + e_k(v), \\[6pt] c^2(v,P) = v(\Gamma+1)\left(P - P_k(v)\right) - v^2 \dfrac{dP_k}{dv}. \end{cases} \qquad (B.1)$$

The parameters $\Gamma$ and $C_v$ represent the Grüneisen coefficient and the specific heat at constant volume respectively. Functions $P_k$ and $e_k$ are specific to the MG-type EOS under consideration. For the JWL they read,

$$\begin{cases} P_k(v) = P_1(v) + k\left(\dfrac{v_{\text{ref}}}{v}\right)^{\Gamma+1}, \\[8pt] e_k(v) = \dfrac{Av_{\text{ref}}}{R_1} e^{-R_1 \frac{v}{v_{\text{ref}}}} + \dfrac{Bv_{\text{ref}}}{R_2} e^{-R_2 \frac{v}{v_{\text{ref}}}} + \dfrac{kv_{\text{ref}}}{\Gamma}\left(\dfrac{v_{\text{ref}}}{v}\right)^{\Gamma} + e_{\text{ref}}, \\[8pt] \dfrac{dP_k}{dv} = -\dfrac{AR_1}{v_{\text{ref}}} e^{-R_1 \frac{v}{v_{\text{ref}}}} - \dfrac{BR_2}{v_{\text{ref}}} e^{-R_2 \frac{v}{v_{\text{ref}}}} - (1+\Gamma)\dfrac{kv_{\text{ref}}}{v^2}\left(\dfrac{v_{\text{ref}}}{v}\right)^{\Gamma}, \\[8pt] P_1(v) = A e^{-R_1 \frac{v}{v_{\text{ref}}}} + B e^{-R_2 \frac{v}{v_{\text{ref}}}}, \\[8pt] k = \left(P_{\text{CJ}} - P_1(v_{\text{CJ}}) - \dfrac{\Gamma C_v T_{\text{CJ}}}{v_{\text{CJ}}}\right)\left(\dfrac{v_{\text{CJ}}}{v_{\text{ref}}}\right)^{\Gamma+1}, \\[8pt] e_{\text{ref}} = e_{\text{CJ}} - \dfrac{Av_{\text{ref}}}{R_1} e^{-R_1 \frac{v_{\text{CJ}}}{v_{\text{ref}}}} - \dfrac{Bv_{\text{ref}}}{R_2} e^{-R_2 \frac{v_{\text{CJ}}}{v_{\text{ref}}}} - v_{\text{CJ}}\dfrac{P_{\text{CJ}} - P_1(v_{\text{CJ}})}{\Gamma}, \\[8pt] v_{\text{CJ}} = v_{\text{ref}} - P_{\text{CJ}}\left(\dfrac{v_{\text{ref}}}{D_{\text{CJ}}}\right)^2, \\[8pt] e_{\text{CJ}} = e_0 + \dfrac{P_{\text{CJ}}}{2}(v_{\text{ref}} - v_{\text{CJ}}), \end{cases} \qquad (B.2)$$

where $A$, $B$, $R_1$, $R_2$, and $e_0$ are fitted parameters depending on the considered material. The subscript "ref" and "CJ" denote the reference state (usually atmospheric) and Chapman-Jouguet (CJ) state respectively. The various thermodynamic parameters of the JWL EOS are adjusted by fitting the adiabat starting from the CJ point. The adiabat is computed with a thermochemical code using the BKW EOS. In the present work, the various JWL parameters for the considered explosives are determined with the Cheetah code using the BKWC library. The parameters are summarized in Table 10.

| Parameter | AN | HMX | NG | NM | PETN | RDX | TATB | TNM | TNT |
|---|---|---|---|---|---|---|---|---|---|
| $\rho_{\text{ref}}$ (kg/m$^3$) | 1050 | 1890 | 1600 | 1130 | 1760 | 1800 | 1850 | 1640 | 1640 |
| $A$ (GPa) | 219.68 | 1195.8 | 816.08 | 225.71 | 912.02 | 1045.7 | 839.89 | 984.84 | 517.53 |
| $B$ (GPa) | 2.72 | 12.16 | 7.59 | 4.65 | 9.51 | 11.27 | 7.51 | 9.50 | 6.20 |
| $R_1$ | 5.65 | 4.77 | 5.05 | 4.67 | 4.75 | 4.78 | 4.77 | 5.87 | 4.72 |
| $R_2$ | 1.42 | 1.08 | 1.04 | 1.10 | 1.08 | 1.12 | 1.10 | 1.56 | 1.07 |
| $\Gamma$ | 0.51 | 0.39 | 0.31 | 0.37 | 0.38 | 0.43 | 0.32 | 0.60 | 0.29 |
| $C_v$ (J/kg/K) | 1837 | 2502 | 2213 | 2674 | 2410 | 2460 | 2418 | 1741 | 2410 |
| $P_{\text{CJ}}$ (GPa) | 4.66 | 38.60 | 22.38 | 11.37 | 30.84 | 34.47 | 27.15 | 13.75 | 19.17 |
| $D_{\text{CJ}}$ (m/s) | 4329 | 9244 | 7701 | 6064 | 8513 | 8920 | 7824 | 6267 | 6843 |
| $T_{\text{CJ}}$ (K) | 1688 | 4124 | 4518 | 3664 | 4360 | 4213 | 3288 | 2652 | 3712 |
| $e_0$ (kJ/kg) | -4562 | 252 | -1640 | -1753 | -1667 | 311 | -562 | 188 | -277 |

Table 10. JWL thermodynamic parameters adjusted with Cheetah using the BKWC library for the various explosives considered.

## Appendix C. Multiphase flow model

In the present contribution, all multiphase computations are done with the homokinetic model of Saurel et al. [13]. This model is a hyperbolic overdetermined system composed of the following equations (presented for two phases $\{1,2\}$ in 1D for the sake of simplicity),

$$\begin{cases} \dfrac{\partial \alpha_1}{\partial t} + u \dfrac{\partial \alpha_1}{\partial x} = \mu(P_1 - P_2), \\ \dfrac{\partial \alpha_1 \rho_1}{\partial t} + \dfrac{\partial \alpha_1 \rho_1 u}{\partial x} = 0, \\ \dfrac{\partial \alpha_2 \rho_2}{\partial t} + \dfrac{\partial \alpha_2 \rho_2 u}{\partial x} = 0, \\ \dfrac{\partial \rho u}{\partial t} + \dfrac{\partial (\rho u^2 + P)}{\partial x} = 0, \\ \dfrac{\partial \alpha_1 \rho_1 e_1}{\partial t} + \dfrac{\partial \alpha_1 \rho_1 e_1 u}{\partial x} + \alpha_1 P_1 \dfrac{\partial u}{\partial x} = -\mu P_I (P_1 - P_2) + h(T_1 - T_2), \\ \dfrac{\partial \alpha_2 \rho_2 e_2}{\partial t} + \dfrac{\partial \alpha_2 \rho_2 e_2 u}{\partial x} + \alpha_2 P_2 \dfrac{\partial u}{\partial x} = \mu P_I (P_1 - P_2) - h(T_1 - T_2), \\ \dfrac{\partial \rho E}{\partial t} + \dfrac{\partial u(\rho E + P)}{\partial x} = 0, \\ \text{with} \quad \mu \to +\infty \text{ and } h \to +\infty \text{ or } 0. \end{cases} \quad (C.1)$$

This flow model is of particular interest in the present context as it is able to deal with two-phase mixtures in mechanical and mechanical-thermal equilibrium, through appropriate relaxation solvers (see [37]). It is also able to deal with material interfaces, as it is capable to match interface conditions of equal pressures and velocities. These two features were intensively used for example in [39] and in a pioneer version by [44].

The notations are the same as before. In addition, the volume fractions $\alpha_k$ of phase $k \in \{1,2\}$ are introduced. The phases are in velocity equilibrium and $u$ denotes the mixture center of mass velocity common to all phases. The mixture density is defined as $\rho = \alpha_1 \rho_1 + \alpha_2 \rho_2$, the mixture pressure is $P = \alpha_1 P_1 + \alpha_2 P_2$, and the mixture total energy is $E = Y_1 e_1 + Y_2 e_2 + u^2/2$, with the mass fraction for each phase $k \in \{1,2\}$ given by $Y_k = \alpha_k \rho_k / \rho$. The interfacial pressure appearing on the right-hand side of (36) reads $P_I = \dfrac{Z_2 P_1 + Z_1 P_2}{Z_1 + Z_2}$, where $Z_k = \rho_k c_k$ denote the acoustic impedance of phase k.

The terms $\mu$ and $h$ represent the pressure and temperature relaxation rates. Stiff pressure relaxation solver provides the relaxed pressure common to all phases and the corresponding densities when pressures reach equilibrium. In addition, following [13], the internal energies are also reset with the pressure at mechanical equilibrium and the relaxed densities with the help of the additional total mixture energy equation and associated mixture EOS. When all phases are governed by the NASG EOS, which is the case in the present work where the thermodynamic relaxation of [10] is used for phases governed by MG-type EOS or BKW EOS, the equilibrium pressure reads,

$$P = \frac{e - \sum Y_k q_k^* - \sum Y_k \left(\dfrac{C_{v,k}^*}{R_k^*} + 1\right)(v_k - b_k^*) P_{\infty,k}^*}{\sum Y_k \dfrac{C_{v,k}^*}{R_k^*}(v_k - b_k^*)}. \tag{C.2}$$

After this step, both phases are in mechanical equilibrium (velocity and pressure). A stiff temperature relaxation solver is also used in the present work between gas detonation products and condensed carbon. More details on stiff relaxation solvers are available for instance in [45,37,10].

System (C.1) is hyperbolic with wave speeds $u$ and $u \pm c$, with the following definition for the mixture square sound speed: $c^2 = Y_1 c_1^2 + Y_2 c_2^2$. Numerical computations have shown excellent convergence to exact solutions for interfaces separating (nearly pure) fluids, even under extreme flow conditions. Details regarding numerical resolution of this model are available for example in [13,46,47].

In the present framework, when the thermodynamic relaxation method of [10] is used for phases governed by the BKW (Section 3.5) or MG-type EOS such as CC EOS, System (C.1) is complemented by transport-relaxation equations for each phase $k \in \{1,2\}$. These equations are written in conservative form with the help of the mixture mass equation,

$$\begin{cases} \dfrac{\partial \rho P^*_{\infty,k}}{\partial t} + \dfrac{\partial \rho P^*_{\infty,k} u}{\partial x} = \dfrac{\rho}{\tau}\left(P_{\infty,k}(v_k,e_k) - P^*_{\infty,k}\right), \\ \dfrac{\partial \rho R^*_k}{\partial t} + \dfrac{\partial \rho R^*_k u}{\partial x} = \dfrac{\rho}{\tau}\left(R_k(v_k,e_k) - R^*_k\right), \\ \dfrac{\partial \rho C^*_{v,k}}{\partial t} + \dfrac{\partial \rho C^*_{v,k} u}{\partial x} = \dfrac{\rho}{\tau}\left(C_{v,k}(v_k,e_k) - C^*_{v,k}\right), \\ \dfrac{\partial \rho q^*_k}{\partial t} + \dfrac{\partial \rho q^*_k u}{\partial x} = \dfrac{\rho}{\tau}\left(q_k(v_k,e_k) - q^*_k\right), \\ \text{with } \tau \to 0^+. \end{cases} \quad (C.3)$$

The implementation of the thermodynamic relaxation method in the multiphase flow model poses no extra difficulty when stiff temperature relaxation is absent. The thermodynamic relaxation step is done independently for each phase after the hyperbolic step and pressure relaxation.

When stiff temperature relaxation is present, the temperature $T_{\text{TEOS}}$ used in the relaxation formulations (21) for the thermodynamic parameters determination must be the equilibrium temperature and not the individual phase temperature. This key point is needed to recover the same results with the conventional method using the Target EOS (TEOS) and the relaxation method using the Predictor EOS (PEOS) when stiff temperature relaxation is present.

## Appendix D. Chapman-Jouguet computation method

In this Appendix, the method used to compute the Chapman-Jouguet (CJ) states given in Section 5.1 is detailed. In the present frame, the conventional CJ theory is adapted to the two-phase mixture made of gas and solid carbon, in both pressure and temperature equilibrium.

The detonation front is treated as a discontinuity and thus obeys the conventional jump conditions of mass, momentum, and energy for the two-phase mixture:

$$\begin{cases} \rho_0(u_0 - D) = \rho(u - D), \\ \rho_0 u_0(u_0 - D) + P_0 = \rho u(u - D) + P, \\ \rho_0 E_0(u_0 - D) + P_0 u_0 = \rho E(u - D) + Pu. \end{cases} \quad (D.1)$$

The subscript "0" indicates the initial state of the unreacted material. $D$ represents the shock velocity, which corresponds to the stationary detonation velocity. At the CJ point, as the detonation is 1D and plane, all the reactive material is burnt, meaning that only detonation products are present. These products contain both gas and solid. System (D.1) is complemented by the CJ relation:

$$D = u + c. \quad (D.2)$$

Moreover, the reactive material is initially at rest, i.e., $u_0 = 0 \text{ m/s}$. The shock-jump relations (D.1) become,

$$\begin{cases} \rho c - \rho_0 D = 0, \\ P - P_0 + \rho c(c - D) = 0, \\ \rho e c - \rho_0 e_0 D + \dfrac{\rho c}{2}(c - D)^2 + P(c - D) = 0. \end{cases} \quad (D.3)$$

At this point, assuming the internal energy and sound speed are known thanks to the detonation products equation of state under the form $e(v,P)$ and $c(v,P)$, the jump relations (D.3) correspond to a non-linear system of three equations with three unknown: $\rho$, $P$ and $D$. When the detonation products are governed by the ideal gas EOS an explicit solution is available (see for example [48]). However, in the present context, detonation products are governed by the BKW-CC EOSs for the solid-gas mixture. Numerical resolution of System (D.3) is consequently addressed.

First, system (D.3) is simplified. The detonation velocity is expressed with the first two equations as,

$$D = \frac{\rho}{\rho_0} c, \qquad D = \frac{P - P_0}{\rho c} + c. \tag{D.4}$$

The relations (D.4) are inserted in (D.3):

$$\begin{cases} \frac{c^2}{v^2}(v - v_0) + P - P_0 = 0, \\ \frac{c^2}{v^2}(e - e_0) - \frac{1}{2}(P^2 - P_0^2) = 0. \end{cases} \tag{D.5}$$

The system (D.5) now has two equations and two unknown: $v$ and $P$. System (D.5) is resolved numerically with a Newton method to obtain the CJ point. In the present work, detonation products are a solid-gas mixture with the gas species (G) governed by BKW EOS and the solid carbon (S) governed by CC. As such, the energy, specific volume, and sound speed present in System (D.5) obey to the following relations,

$$e = Y_S e_S(P,T) + Y_G e_G(P,T,y_i), \tag{D.6}$$

$$v = Y_S v_S(P,T) + Y_G v_G(P,T,y_i), \tag{D.7}$$

$$c^2 = -v^2 \frac{C_p}{C_v} \left.\frac{\partial P}{\partial v}\right|_T,$$

$$\text{with} \quad C_v = \left.\frac{\partial e}{\partial T}\right|_v = Y_S \left.\frac{\partial e_S}{\partial T}\right|_{v_S} + Y_G \left.\frac{\partial e_G}{\partial T}\right|_{v_G, y_i},$$

$$C_p = \left.\frac{\partial h}{\partial T}\right|_v = Y_S \left.\frac{\partial h_S}{\partial T}\right|_{v_S} + Y_G \left.\frac{\partial h_G}{\partial T}\right|_{v_G, y_i}, \tag{D.8}$$

$$\left.\frac{\partial P}{\partial v}\right|_T = \left(\left.\frac{\partial v}{\partial P}\right|_T\right)^{-1} = \left(Y_S \left(\left.\frac{\partial P}{\partial v_S}\right|_T\right)^{-1} + Y_G \left(\left.\frac{\partial P}{\partial v_G}\right|_{T,y_i}\right)^{-1}\right)^{-1}.$$

Where $h = e + Pv$ is the specific enthalpy. $C_v$ and $C_p$ are respectively the specific heat at constant volume and constant pressure of the mixture. For the CJ computations in this work, the composition of the detonation products is computed beforehand with Cheetah. So, the mass fractions $Y_S$, $Y_G = 1 - Y_S$, and $y_i = Y_i / Y_G$ used in BKW, are known. The various thermodynamic functions involved in relations (D.6), (D.7) and (D.8) are given for the gas phase BKW in Section 2 and for the solid phase in Section 4. Finally, System (D.5) complemented by the mixture relations (D.6), (D.7) and (D.8) is a non-linear system with three unknown: $P$, $v$ and $T$. An approximate two-variables Newton method is used to determine $P$ and $v$ with functions (D.5), where the mixture energy and sound speed are given by (D.6) and (D.8) respectively. The method is approximated in the sense that the functions derivatives are determined by numerical differentiation. During resolution, when the temperature $T$ is needed, its determination is addressed with another Newton method resolving relation (D.7) for the current specific volume $v$.

## CRediT authorship contribution statement

**Loann Neron:** Conceptualization, Methodology, Software, Formal analysis, Investigation, Data Curation, Writing – Original Draft, Writing – Review & Editing, Visualization. **Richard Saurel:** Conceptualization, Methodology, Formal analysis, Investigation, Resources, Writing – Original Draft, Writing – Review & Editing, Supervision, Project administration.

## Declaration of competing interest

The authors declare that they have no known competing financial interests or personal relationships that could have appeared to influence the work reported in this paper.

## References

[1] Lee, E.L., Horning, H.C. & Kury, J.W., "Adiabatic Expansion of High Explosives Detonation Products", Lawrence Radiation Laboratory, University of California, Livermore, TID 4500-UCRL 50422 (1968)

[2] Fried, L. E., Howard, W. M., & Souers, P. C., "Cheetah 2.0 user's manual", UCRL-MA-117541 Rev. 5. Lawrence Livermore National Laboratory, August (1998)

[3] Suceska, M., "Evaluation of detonation energy from EXPLO5 computer code results", Propellants, Explosives, Pyrotechnics, 24(5), 280-285 (1999)

[4] Arnold, W., Rottenkolber, E., & Hartmann, T., "DRAGON-The German Thermo-Chemical Code Based on the Becker-Kistiakowsky-Wilson Equation of State", Propellants, Explosives, Pyrotechnics, 48(1) (2023)

[5] Souers, P. C., Anderson, S., Mercer, J., McGuire, E., & Vitello, P., "JWL++: A simple reactive flow code package for detonation", Propellants, Explosives, Pyrotechnics, 25(2), 54-58 (2000)

[6] Chinnayya, A., Daniel, E., & Saurel, R., "Modelling detonation waves in heterogeneous energetic materials", Journal of Computational Physics, 196(2), 490-538 (2004)

[7] Saurel, R., Fraysse, F., Furfaro, D., & Lapebie, E., "Reprint of: Multiscale multiphase modeling of detonations in condensed energetic materials", Computers & Fluids, 169, 213-229 (2018)

[8] Hobbs, M. L., & Baer, M. R., "Calibrating the BKW-EOS with a large product species data base and measured CJ properties", In 10th International Symposium on Detonation, Boston, MA (1993)

[9] Suceska, M., Chan, H. Y., & How-Ghee, A., "Can the accuracy of BKW EOS be improved?", In 15th International Detonation Symposium Proceedings, vol. 6, San Francisco, USA, 1247-1256 (2014)

[10] Neron, L., Saurel, R., Chiapolino, A., & Fraysse, F., "Relaxation method for Mie-Grüneisen type equation of state", Physics of Fluids, 35(11) (2023)

[11] Cochran, S. G. & Chan, J., "Shock initiation and detonation models in one and two dimensions", UCID-18024. California Univ., Livermore (1979)

[12] Suceska, M., Dobrilovic, M., Bohanek, V., & Stimac, B., "Estimation of explosive energy output by EXPLO5 thermochemical code." Zeitschrift für anorganische und allgemeine Chemie, 647(4), 231-238 (2021)

[13] Saurel, R., Petitpas, F., & Berry, R. A., "Simple and efficient relaxation methods for interfaces separating compressible fluids, cavitating flows and shocks in multiphase mixtures", Journal of Computational Physics, 228(5), 1678-1712 (2009)

[14] Mader, C. L., "Detonation properties of condensed explosives computed using the Becker-Kistiakowsky-Wilson equation of state", Los Alamos Scientific Laboratory, University of California (1963)

[15] Suceska, M., Chan, H. Y., Stimac, B., & Dobrilovic, M., "BKW EOS: History of Modifications and Further Improvement of Accuracy with Temperature-Dependent Covolumes of Polar Molecules", Propellants, Explosives, Pyrotechnics, 48(1), e202100278 (2023)

[16] Becker, R., "Stosswelle und detonation", Zeitschrift für Physik, 8(1), 321-362 (1922)

[17] Kistiakowsky, G. B., & Wilson, E. B., "The hydrodynamic theory of detonation and shock waves", Office of the Publication Board, Department of Commerce (1941)

[18] Cowan, R. D., & Fickett, W., "Calculation of the detonation properties of solid explosives with the Kistiakowsky-Wilson equation of state", The Journal of Chemical Physics, 24(5), 932-939 (1956)

[19] Kee, R. J., Rupley, F. M., Miller, J. A., & Jefferson, T. H., "CHEMKIN: A general-purpose, problem-independent, transportable, FORTRAN chemical kinetics code package", Sandia Report SAND-80-8003, Sandia National Laboratory, Albuquerque, NM (1980)

[20] Callen, H. B., & Kestin, J., "An Introduction to the Physical Theories of Equilibrium Thermostatics and Irreversible Thermodynamics", Wiley, New York (1960)

[21] Finger, M., Lee, E., Helm, F. H., Hayes, B., Hornig, H., McGuire, R., Kahara, M., & Guidry, M., "Effect of elemental composition on detonation behavior of explosives", Lawrence Livermore Laboratory, University of California, Livermore (USA) (1976)

[22] Fried, L. E., & Souers, P., "BKWC: An empirical BKW parametrization based on cylinder test data", Propellants, Explosives, Pyrotechnics, 21(4), 215-223 (1996)

[23] Suceska, M., Ang, H. G., & Chan, H. Y., "Modification of BKW EOS introducing density-dependent molecular covolumes concept", In Materials Science Forum, vol. 673, 47-52 (2011)

[24] Chase, M. W., "NIST-JANAF thermochemical tables", J. Phys. Chem. Ref. Data (1998)

[25] Mader, C. L., "FORTRAN BKW: A code for computing the detonation properties of explosives", Report No. LA-3704, Los Alamos Scientific Laboratory, Los Alamos, NM (1967)

[26] Butler, P. B., "Real gas equations of state for CHEMKIN", Sandia Report SAND88-3118, Sandia National Laboratory, Albuquerque, NM (1989)

[27] Fried, L. E., "Cheetah 1.0 user's manual", Lawrence Livermore National Laboratory, Livermore, CA (1994)

[28] Toro, E. F., Spruce, M., & Speares, W., "Restoration of the contact surface in the HLL-Riemann solver", Shock waves, 4, 25-34 (1994)

[29] Toro, E. F., "Riemann solvers and numerical methods for fluid dynamics: A practical introduction", Springer Science & Business Media (1999)

[30] Saurel, R., Huber, G., Jourdan, G., Lapébie, E., & Munier, L. "Modelling spherical explosions with turbulent mixing and post-detonation." Physics of Fluids, 24(11) (2012)

[31] Le Métayer, O., & Saurel, R., "The Noble-Abel stiffened-gas equation of state", Physics of Fluids, 28(4), 046102 (2016)

[32] Howard, W. M., Fried, L. E., & Souers, P. C., "Kinetic modeling of non-ideal explosives with CHEETAH", In 11th International Symposium on Detonation, Snowmass, Colorado (1998)

[33] Liu, Q., Duan, Y., Ma, H., Long, X., & Han, Y., "Review on the exploration of condensed carbon formation mechanism in detonation products", AIP Advances, 10(5) (2020)

[34] Bushman, A.V., Lomonosov, I.V. & Khishchenko, K. V., "Shock wave data base", http://www.ficp.ac.ru/rusbank/ (2004)

[35] Heuzé, O., & Bauer, P., "A simple method for the calculation of the detonation properties of CHNO explosives", In Symposium HDP, La Grande Motte, France, 225-232 (1989)

[36] Fried, L. E., & Souers, P., "Cheetah: A next generation thermochemical code", Report No. UCRL-ID-117240, Lawrence Livermore National Laboratory, Livermore, CA (1994)

[37] Le Métayer, O., Massoni, J., & Saurel, R., "Dynamic relaxation processes in compressible multiphase flows. Application to evaporation phenomena", In Esaim: Proceedings (Vol. 40, pp. 103-123). EDP Sciences (2013)

[38] Saurel, R., Petipas, F., & Abgrall, R., "Modelling phase transition in metastable liquids: application to cavitating and flashing flows", Journal of Fluid Mechanics, 607, 313-350 (2008)

[39] Petitpas, F., Saurel, R., Franquet, E., & Chinnayya, A., "Modelling detonation waves in condensed energetic materials: Multiphase CJ conditions and multidimensional computations", Shock waves, 19, 377-401 (2009)

[40] Kamlet, M. J., & Jacobs, S. J. "Chemistry of detonations. I. A simple method for calculating detonation properties of C–H–N–O explosives", The Journal of Chemical Physics, 48(1), 23-35 (1968)

[41] Godunov, S. K., Zabrodin, A. V., Ivanov, M. I., Kraiko, A. N., & Prokopov, G. P., "Numerical solution of multidimensional problems of gas dynamics", Moscow Izdatel Nauka (1976)

[42] Menikoff, R., & Plohr, B. J., "The Riemann problem for fluid flow of real materials", Reviews of modern physics, 61(1), 75 (1989)

[43] Neron, L., & Saurel, R., "Noble–Abel/first-order virial equations of state for gas mixtures resulting from multiple condensed reactive materials combustion", Physics of Fluids, 34(1), 016107 (2022)

[44] Abgrall, R., Nkonga, B., & Saurel, R. "Efficient numerical approximation of compressible multi-material flow for unstructured meshes." Computers & Fluids, 32(4), 571-605 (2003)

[45] Lallemand, M. H., Chinnayya, A., & Le Métayer, O., "Pressure relaxation procedures for multiphase compressible flows", International Journal for Numerical Methods, 49(1), 1-56 (2005)

[46] Chiapolino, A., Saurel, R., & Nkonga, B., "Sharpening diffuse interfaces with compressible fluids on unstructured meshes", Journal of Computational Physics, 340, 389-417 (2017)

[47] Saurel, R., & Pantano, C., "Diffuse-interface capturing methods for compressible two-phase flows", Annual Review of Fluid Mechanics, 50, 105-130 (2018)

[48] Fickett, W., & Davis, W. C., "Detonation: theory and experiment", University of California Press., Berkeley (1979)